\documentclass[aip,jcp,
 amsmath,amssymb,
 reprint,floatfix
]{revtex4-2}
\pdfoutput=1
\usepackage[dvipsnames]{xcolor}

\usepackage{graphicx}
\usepackage{amsmath}
\usepackage{dcolumn}
\usepackage{bm}
\usepackage[utf8]{inputenc}
\usepackage[T1]{fontenc}
\usepackage{mathptmx}
\usepackage{comment}
\usepackage{float}
\usepackage{placeins}
\usepackage{xr-hyper}
\usepackage{hyperref}
\DeclareGraphicsExtensions{.pdf}
\newcommand{\markup}[1]{{\color{black} #1}}

\newcommand{\G}[1]{\Delta G_{\textrm{#1}}}
\newcommand{\ts}[1]{\langle \tau_{\textrm{#1}} \rangle}
\newcommand{\Meff}{[\textrm{M}_{\textrm{eff}}]}
\newcommand{\M}{[\textrm{M}]}
\newcommand{\pcomp}{p_{\textrm{complete}}}
\newcommand{\avg}[1]{\langle #1 \rangle}
\newcommand{\etal}{{\it et al.} }

\graphicspath{ {./Figures/} }
\begin{document}
\preprint{AIP/123-QED}
\title[]{Minimal mechanism for cyclic templating of length-controlled copolymers under isothermal conditions}
\author{Jordan Juritz}
\affiliation{Department of Bioengineering and Centre for Synthetic Biology, Imperial College London, London SW7 2AZ, United Kingdom}
\author{Jenny M. Poulton}
\affiliation{Foundation for Fundamental Research on Matter (FOM) Institute for Atomic and Molecular Physics (AMOLF), 1098 XE Amsterdam, The Netherlands}%
\author{Thomas E. Ouldridge}
\email{t.ouldridge@imperial.ac.uk}
\affiliation{Department of Bioengineering and Centre for Synthetic Biology, Imperial College London, London SW7 2AZ, United Kingdom}
\date{\today}

\begin{abstract}
The production of sequence-specific copolymers using copolymer templates is fundamental to the synthesis of complex biological molecules and is a promising framework for the synthesis of synthetic chemical complexes. Unlike the superficially similar process of self-assembly, however, the development of synthetic systems that implement templated copying of copolymers \markup{ under constant environmental conditions} has been challenging. The main difficulty has been overcoming product inhibition, or the tendency of products to adhere strongly to their templates -- an effect that gets exponentially stronger with template length. \markup{ We develop coarse-grained  models of copolymerisation on a finite-length template and analyse them through stochastic simulation}. We use these models first to demonstrate that product inhibition prevents reliable template copying, and then ask how this problem can be overcome to achieve cyclic production of polymer copies of the right length and sequence in an autonomous and chemically-driven context. We find that a simple addition to the model is sufficient to generate far longer polymer products that  initially form on, and then separate from, the template. In this approach, some of the free energy of polymerisation is diverted into disrupting copy-template bonds behind the leading edge of the growing copy copolymer. By additionally weakening the final copy-template bond at the end of the template, the model predicts that reliable copying with a high yield of full-length, sequence-matched products is possible over large ranges of parameter space, opening the way to the engineering of synthetic copying systems \markup{ that operate autonomously.}
\end{abstract}

\maketitle

\section{Introduction}

\markup{ Copolymers -- polymers formed from two or more types of monomer unit -- are ubiquitous in biology. DNA, RNA and proteins are biological copolymers for which the information pertaining to the function and/or the structure of the molecule is encoded into the sequence of the copolymer\cite{CRICK1970}. There is a huge diversity of copolymer sequences to be found in biology; the human proteome, for instance, is comprised of roughly 25,000 different proteins, each consisting of linear chains assembled from the 20 amino acid residues \cite{Stumpf2008}, with a median length of 375 amino acids \cite{Brocchieri2005}.

How are these complex molecules formed? In general, it is impossible to encode and consistently assemble thousands of distinct and essentially arbitrary macromolecules, each with approximately $ 375$ units, through the self-assembly of just 20 types of building blocks\cite{Sartori2020}. Instead, these copolymers are produced by copying copolymer templates, such as DNA and mRNA. In templated copying processes, complementary interactions between monomers and the template direct the assembly of the product so that distinct, arbitrary copolymer sequences can be reliably created from a common set of monomers \cite{Cabello-Garcia2021}. In extant organisms, templated copying is supported by a large amount of the cellular resources and is aided by ensembles of complex molecular machines, such as DNA polymerases involved in DNA replication \cite{Reha-Krantz2010,Joyce2004}, RNA polymerases in transcription \cite{Vannini2013,Ebright2000,Hirose2000}, and ribosomes in translation \cite{Rodnina2018,Fox2010,Petrov2015,Bowman2020,Nikolay2015,Caetano-Anolles2015}. The physics of templated copolymer copying therefore lies at the heart of the synthesis of the diverse and complex molecules of biology \cite{CRICK1970}. 

While recent decades have seen the engineering of remarkably complex self-assembling systems made of DNA \cite{Seeman1982,Rothemund2006,Lin2006,Ke2012,Seeman2017}, proteins \cite{Divine2021, Shen2018}, and combinations of DNA and proteins \cite{Jin2019}, advances in the field of synthetic templating have been comparatively slow. This disparity suggests a poor understanding of templating relative to self-assembly, and represents a missed opportunity to harness biology's most important mechanism of producing chemical complexity - both to generate synthetic sequence-controlled polymers \cite{Lutz2013} and in the context of combinatorial molecular discovery \cite{Usanov2018,OReilly2017}. In this work, we simulate coarse-grained models to investigate a class of templated copolymerisation reactions that may enable enzyme-free copying of templates analogously to the copying processes found in biology. 

We look to biological systems to set the requirements and conditions of operation for the non-enzymatic, synthetic template-copying system we wish to engineer. First, biological copying systems are capable of the accurate reproduction of copolymer sequences from arbitrary but specific lengths of copolymer template, whether by replication \cite{Reha-Krantz2010,Joyce2004}, transcription \cite{Vannini2013,Ebright2000,Hirose2000}, or translation \cite{Rodnina2018,Fox2010,Petrov2015,Bowman2020,Nikolay2015,Caetano-Anolles2015}.  Secondly, each template can be reused many times to produce many copies; the copying process net generates complex molecules that persist separately from their template\cite{Ouldridge2017, Cabello-Garcia2021, Poulton2019, Poulton2021}. This fact is crucial, since templates themselves are necessarily complex molecules \cite{Ouldridge2017, Cabello-Garcia2021, Poulton2019, Poulton2021}. If each new copy consumes -- or remains bound to -- a template, then a new template must be generated from scratch for each copy formed. Thirdly, natural copying systems are capable of operating in spatio-temporally constant environmental conditions, without perturbations, exploitable spatial gradients or other external interventions; they are driven solely by the chemical free energy of the dissolved building blocks and "fuel" molecules, such as ATP \cite{MolBiolCell}. We describe copying that does not rely on such external factors as {\em autonomous}.

The phenomenon of "product inhibition" presents a major challenge to achieving reliable, repeatable production of specific copolymers by templating under autonomous conditions\cite{Michaelis2014}. As monomers polymerise on the template, the product's length increases and the binding free energy between product and template typically grows linearly. Therefore, the likelihood that a product spontaneously detaches from the template is exponentially suppressed with length, inhibiting subsequent copying of the template. As we explore in Section \ref{subsec:basic}, products of varying -- but short -- lengths result when copying long templates in such a setting. Due to product inhibition, long copies may fail to be released from the template, sequestering the template and slowing the copying rate, while short, fragmented copies may be released rapidly. Therefore product inhibition poses significant challenges for the reliability of both length-control and template reuse.

Michaelis \etal \cite{Michaelis2014} emphasised that, in non-enzymatic systems restricted to isothermal operation, strategies that reduce the affinity between products and the template are required to achieve high turnovers (defined as the final ratio of products to templates), even for dimeric templates\cite{Abe2004}. Enforcing separation by reducing the affinity between the copy and template, without compromising the template's reuse, has also proved challenging. Osuna Gálvez and Bode recently reported a templated reaction in which the dimerisation of the reactants and the disruption of {\it both} the reactant-template bonds occurred simultaneously, drastically reducing the product-template affinity\cite{OsunaGalvez2019}. However, this came at the cost of scarring the template, which was was unable to promote further reactions \cite{OsunaGalvez2019}.

Instead, time-varying, non-autonomous environmental conditions are commonly used to drive cyclic templating and product release. In polymerase chain reactions, for instance, cycles of heating dissociate the product from the template and enable further rounds of amplification\cite{Braun2004}. Various other time-varying strategies have also been used to drive enzyme-free, dimeric \cite{He2017,Zhuo2019} and longer template-copying and replicating systems \cite{Nunez-Villanueva2019,Nunez-Villanueva2019Replicate,Nunez-Villanueva2021Accounts,Nunez-Villanueva2021Mutation,Kuhnlein2021}. 

Others have employed non-chemical energy or exploited spatial gradients to engineer systems that spontaneously separate copies from the template, thereby favouring the cyclic assembly and separation of longer copies \cite{Schulman2012,Braun2004,Mast2010,Mast2013,Kreysing2015,Colomb-Delsuc2015}. In Schulman's replicator, shear flow was used to fragment the layers of information-bearing DNA tile-based crystals, revealing more reusable templating surfaces, which enabled exponential self-replication of the crystal\cite{Schulman2012}. Braun and colleagues have exploited the convective currents generated by thermal gradients, inspired by oceanic thermal vents \cite{Baross1985}, with templating and polymerisation occurring in cool regions and separation in hot regions\cite{Braun2004,Mast2010,Mast2013}. Similar environments with spatially and/or temporally varying conditions are thought to have played an important role in the emergence of life \cite{Orgel1968,Crick1968,Gilbert1986,Joyce1989,Mutschler2015,Orgel2004,Szostak2012,Kudella2021, Rosenberger2021}, though these conditions are not required for the operation of modern biological copying enzymes. 

Recently, a novel DNA strand displacement motif, handhold-mediated strand displacement, was used to autonomously drive dimer formation directed by templates \cite{Cabello-Garcia2021}. Here, binding between monomers on a template weakens the connection of one monomer to the template, limiting product inhibition. In principle, the mechanism would allow efficient dimerisation without template scarring. However, it is unclear whether such a mechanisms can scale to overcome product inhibition and reliably produce length- and sequence-controlled copies on longer templates. 

From a theoretical perspective, many authors have considered the permanent deposition of a copy on a template \cite{Bennett1979,Ehrenberg1980,Cady2009,Andrieux2008,Gaspard2014,Sartori2013,Sartori2015,Esposito2010}, though effects of subsequent separation has not been a focus until recently. Moreover, while these previous works have considered mechanisms of sequence-control without copy separation, mechanisms for precise length-control as observed in nature have been neglected. 

More recently, models of templated copolymerisation have been used to study self-replicating molecules under prebiotically plausible conditions \cite{Rosenberger2021,Tupper2021}. Tupper and Higgs have argued that a rolling-circle mechanism could have overcome product inhibition to promote non-enzymatic RNA replication in an RNA-world \cite{Tupper2021}. In this high-level model, the chemical details of the directional polymerisation mechanism were not considered, and nor was precise length control. In a recent theoretical and experimental work on pre-enzymatic templated assembly, Rosenberger \etal demonstrated that polymer aggregates with increased length can be generated from short building blocks under isothermal conditions, though, these polymers were bound up in complexes, not separated from their templates \cite{Rosenberger2021}. Moreover, observing the gradual lengthening of copolymers on average was the priority, not precise length control, as required for perfect information copying. 

The thermodynamic constraints that copy-template separation place on sequence-control were considered in Ref.~\cite{Ouldridge2017}, though non-chemical means were invoked to separate template and copy. In Refs.~\cite{Poulton2019, Poulton2021}, the thermodynamic and kinetic consequences of an isothermal mechanism for generating sequence-controlled copolymers with product separation were considered, but the actual mechanistic details were implicit. Nonetheless, a key theoretical result of these works is that separating an accurate copy from its template necessitates producing a state that is extremely far from equilibrium. This thermodynamic argument also applies to length control: in the absence of residual interactions with a template, an ensemble of copolymers of a specific (but arbitrary and template-selected) length is extraordinarily far from the equilibrium of a broad, exponential distribution of lengths. It is this fundamental physical principle that is practically manifest as the challenge of product inhibition: how can interactions be tuned to allow templates to act as a catalyst for the production of a specific, far-from-equilibrium product state, while avoiding a stable equilibrium of copy-template complexes? 

In this paper we argue that decoupling the length of a copy from its affinity for the template, by the channelling the free energy released in polymerisation to the disruption of copy-template bonds as achieved by HMSD \cite{Cabello-Garcia2021}, would be sufficient to overcome product inhibition even on long templates in an isothermal environment. Further, we hypothesise that additionally weakening the copy-template connection at the final template site would result in the selective release of complete copies rather than shorter fragments. We explore these arguments through simulations of coarse-grained models.

In Section \ref{sec:model} we introduce a basic model of isothermal templated copolymerisation with separation. In Section \ref{subsec:basic}, we demonstrate that cooperative binding to the template prohibits the release of long polymers under a basic growth mechanism, demonstrating that it is necessary to have more complex interactions between copy and template in order to promote reliable separation in a constant environment. In Section \ref{subsec:generic}, we propose and investigate a mechanism which could alleviate the cooperative effect of product inhibition resulting in an increase in the mean polymer length. Extending this mechanism in Section \ref{subsec:weakend}, we bias the production of complete polymers by selectively weakening the final site on the template. Finally, in Section \ref{subsec:accuracy}, we demonstrate that long and accurate copies of the template can be generated if the correct and incorrect monomers have varying binding rates.
}
\section{Model and methods}

\subsection{Model}
\label{sec:model}

\subsubsection{State space}

\begin{figure}[h]
	\centering
	\includegraphics[width=0.8\columnwidth]{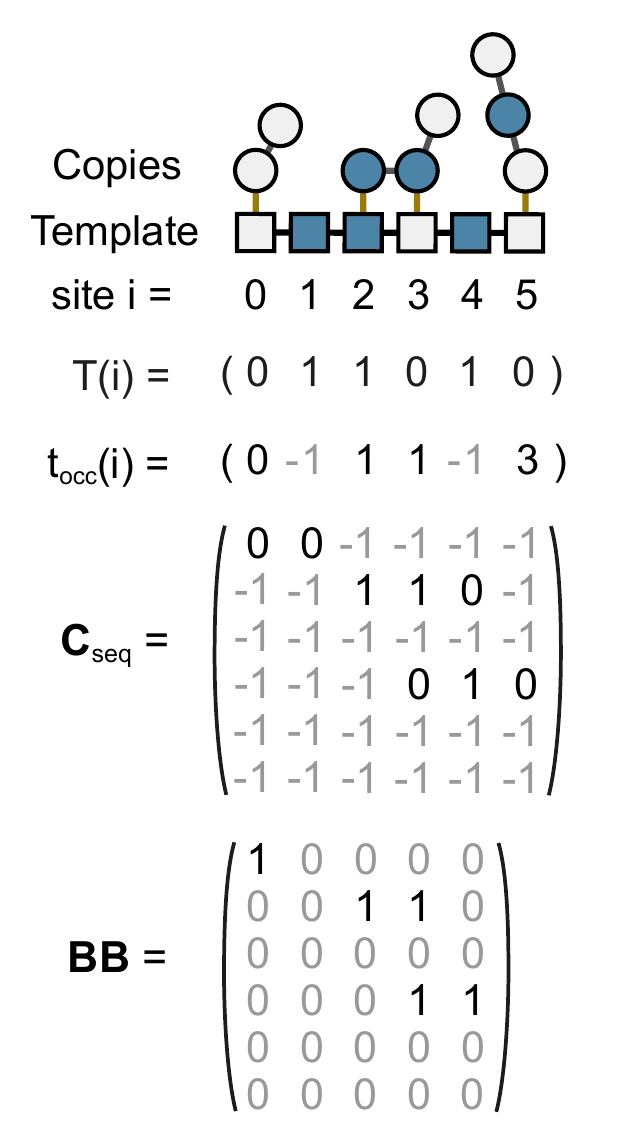}
	\caption{The state of the system is described by $t_{\textrm{occ}}$, $\mathbf{C}_{\textrm{seq}}$ and $\mathbf{BB}$. Sites on the template are indexed with i. The array $T(i)$ stores the template sequence. $t_{\textrm{occ}}$ is an array containing the polymer label of each copy unit attached to the template. The label is the row of  $\mathbf{C}_{\textrm{seq}}$ in which the polymer's sequence is stored. $\mathbf{BB}$ stores the backbone bonds. In this example, three polymers (labelled 0, 1 and 3) are attached by one, two and one bond/s to the template, respectively.}
	\label{fig:DataStructure}
\end{figure}

As shown in Fig.~\ref{fig:DataStructure}, we consider a copolymer template $T = t_{1}t_{2}t_{3}...t_{L_{T}}$, where $t_{i}$ is an integer indicating the monomer type that can take values ${0,1,...,\alpha_{T}-1}$. $\alpha_{T}$ is the number of  distinct types of monomeric unit in the template. A single template copolymer is suspended in a large-volume, well-mixed bath of a second distinct type of monomers that interact  with the template. The identity of these monomers is labelled by an integer that can take values ${0,1,...,\alpha_{C}-1}$.  There are $\alpha_{C}$ distinct types of copy monomer, the $i^{th}$ of which has a concentration $\M_{i}$. These monomers can bind onto the template at any unoccupied site on the template. In this paper, we consider both homogeneous copy-template systems in which there is a single monomer type, $\alpha_{T} = \alpha_{C} = 1$, and also information-bearing binary systems in which there are two monomer types for each of the copy and template monomers $\alpha_{T}=\alpha_{C}=2$. Following Ref.~\cite{Poulton2019}, we assume that copy/template interactions are symmetric with respect to interchange of 0 and 1, giving identical dynamics for all template sequences. We shall use the uniform template $T=000...0$ in all simulations. When the copy-template alphabets are binary, $\alpha_{T}=\alpha_{C}=2$, monomers of type 0 and 1 in the copy copolymer can then simply be interpreted as ``correct" and ``incorrect" matches, respectively. We will refer to all molecules as polymers unless their nature as copolymers is important.

A copy unit with no neighbouring backbone bonds is a monomer. As shown in Fig.~\ref{fig:DataStructure}, copy polymers are chains of copy units that share a `backbone' linking one unit to the next. Each copy monomer may form up to two backbone bonds, one with a copy unit ahead and one with a copy unit behind with the boundary condition that backbone bonds can only be formed ahead at site $0$, and behind at site $L_{T}-1$. As can be seen in Fig.~\ref{fig:DataStructure}, copy polymers can occupy contiguous stretches of template sites and may have `tails', portions of their length which are not directly bound to the template, but are indirectly tethered to the template by bound units with which they share a backbone. These tails may be present at either end of the polymer. Copy polymer `bridges', unbound stretches between bound units within the same copy polymer, are neglected. It is assumed that such behaviour is strongly suppressed due to the constrained close proximity enforced by the bound units, just as internal bubbles are strongly suppressed between complementary nucleic acid duplexes \cite{SantaLucia2004}.

We assume that the concentration of the monomer species $[M_{i}]$ in the bath is buffered to be constant. In the simulations that follow, we will initiate the system with no polymers in solution, and with the template empty (all sites unoccupied). Under these conditions, and, again, given that the volume of the bath is large, we further assume that the concentration of copy polymers in the bath remains at a negligible level compared to the concentrations of copy monomer. That is to say, once polymers fully unbind from the template, they diffuse into the bath and do not bind to the template again. We therefore neglect possible product inhibition due to very high concentrations of products, a secondary problem relative to product inhibition due to products that simply never detach from the template. In the models presented here, polymers cannot bind to the template from solution and bridges cannot form loops that cause the contour length of a copy polymer to exceed the length of the template, $L_{T}$, and, therefore, only copy polymers with length $L \leq L_{T}$ can be produced. 

The state of the system is thus uniquely specified by the sequence of monomers in all growing copy polymers, including those with only one unit, and by specifying which site each monomer is attached to on the template (if it is attached to no template site, the monomer exists in a tail) as outlined in Fig.~\ref{fig:DataStructure}. We do not consider the conformations of the system within each of these macrostates.

\begin{figure}[t]
	\centering
	\includegraphics[width=\columnwidth]{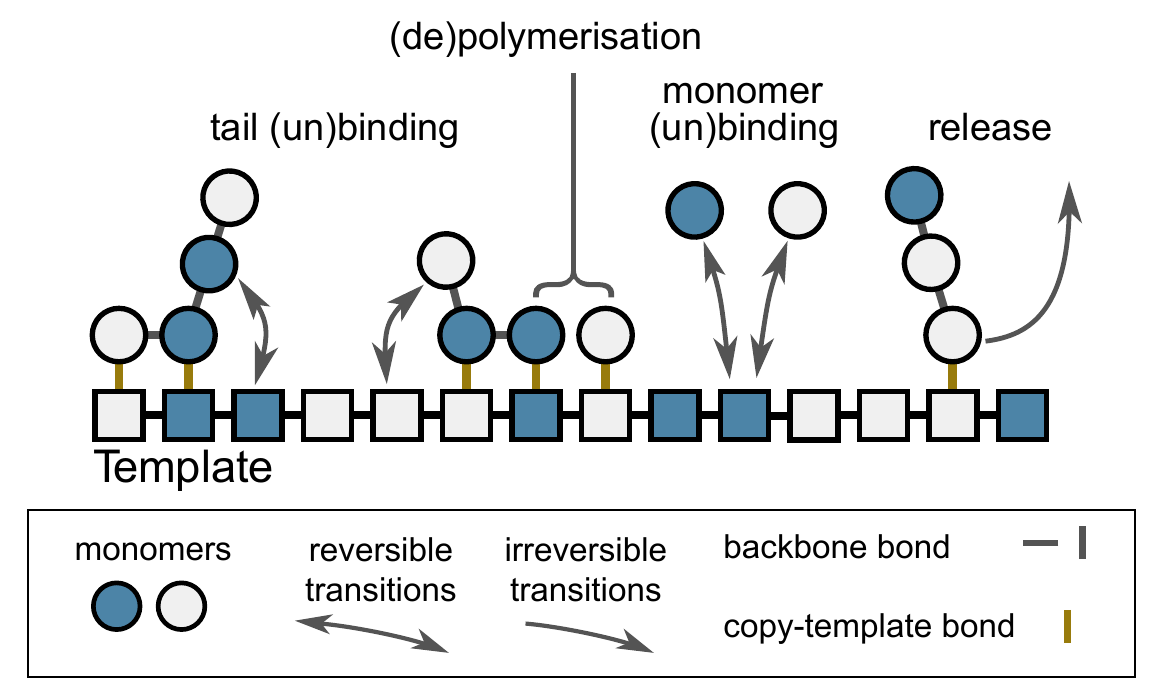}
	\caption{Simple transition rules between coarse grained states of a templated polymerisation process. Each site on the template can be occupied by one monomeric unit. Monomers may bind to and unbind from the template from solution and from the tails of copolymers. Neighbouring units on the template may polymerise or depolymerise. All transitions, aside from the final release of a polymer from the template, are reversible, and can occur at any site on the template.}
	\label{fig:ModelAnywhere}
\end{figure}

As we indicate in Fig.~\ref{fig:DataStructure}, sites in the system are indexed with $i$ running from $0$ to $L_{T}-1$. The $i^{th}$ site on the template can be unoccupied or occupied by a copy unit; either a monomer or part of a polymer. Each isolated monomer or polymer on the template is given a unique label. If two copy units share a backbone, even indirectly via other monomers, they must be part of the same polymer, and therefore they share the same label. The template can be occupied by at most $L_{T}$ distinct monomers or polymers at any moment, hence polymer labels run from $0$ to $L_{T}-1$. We introduce this label to enable easy accounting when extending, fragmenting or recombining polymers. The template occupation labels are stored in a vector $t_{\textrm{occ}}$, where the $i^{th}$ component, $t_{\textrm{occ}}(i)$, is the label $l$ of the copy unit bound to the template at site $i$, or $-1$ if the site is empty. 

The sequence of all polymers - including units in tails- and monomers in the system is stored in an $L_{T} \times L_{T}$ matrix, $\mathbf{C}_{\textrm{seq}}$, in which the $l^{th}$ row contains the sequence of the copy unit with label $l$, buffered with -1s before and after the sequence. Therefore, $\mathbf{C}_{\textrm{seq}}(t_{\textrm{occ}}(i),i)$ is the identity of the copy unit occupying the template at site $i$. 

Each copy unit may form up to two backbone bonds, one with a copy unit ahead and one with a copy unit behind, with the boundary condition that backbone bonds can only be formed ahead at site $0$, and behind at site $L_{T}-1$. The $L_{T} \times L_{T}-1$ matrix $\mathbf{BB}$  stores the state of all backbone bonds in the system. The $i^{th}$ column of the $l^{th}$ row of $\mathbf{BB}$ is $1$ if there is a backbone bond between copy units with label $l$ at sites $i$ and $i+1$ (not necessarily connected to the template), and $0$ otherwise.

\subsubsection{Transition rules}
\label{sec:transitions}

\begin{figure*}[t]
	\centering
	\includegraphics[width=\textwidth]{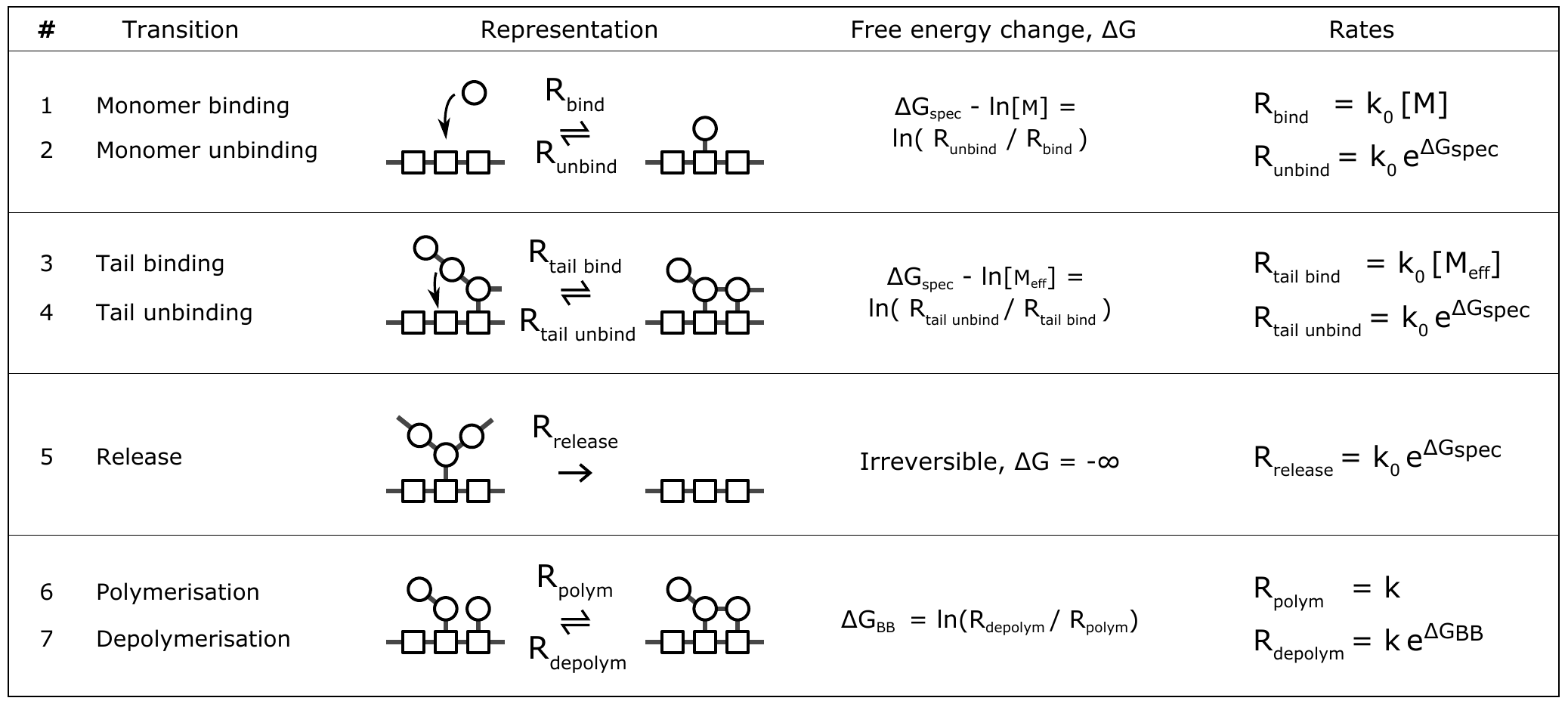}
	\caption{Seven types of transition constitute the basic model of templated polymerisation. Each transition modifies the system at a single site, or, in the case of polymerisation, between neighbouring sites. The free energy change associated with the transition constrains the log of the ratio of backwards to forwards transitions according to the principle of local detailed balance, which reduces the number of free parameters in our model and assures that the model is thermodynamically self-consistent. Final release of a polymer is assumed to be irreversible, and hence is associated with an infinite free energy change.}
	\label{fig:BasicModel}
\end{figure*}

The transition rules that define the permitted state changes are listed in Fig.~\ref{fig:BasicModel} and illustrated in in Fig.~\ref{fig:ModelAnywhere}. All transitions in this model are microscopically reversible, apart from the final detachment of copolymers, since the concentration of products in solution, and hence the rate of polymer rebinding, is assumed to be zero. 
The transitions that feature in this basic model are as follows:
\vspace{2mm}
\begin{enumerate}
\item A copy monomer in solution may diffuse close enough to an unoccupied site $i$ on the template for an attractive interaction to cause it to bind there. 
\item The reverse reaction may also occur, in which a monomer bound at site $i$ on the template may unbind and freely diffuse away from the template.
\item  A unit of a copy polymer that is either bound at the end of a polymer or is the last bound unit before a tail may unbind from the template. This increases the length of the tail by $1$. 
\item In reverse, the first unit in the tail of a polymer hovering over site $i$ (which must have a template bound and backbone-linked neighbour at either site $i-1$ or site $i+1$) may bind to site $i$ if the site is not already occupied.
\item If the last bond between a polymer and the template breaks, the polymer is released. This liberated polymer mixes with the large volume and never returns to bind to the template. 
\item Polymerisation can only occur between available, template-bound neighbours. Specifically, if a bound copy unit at site $i$ can make a forward backbone bond to site $i+1$, and a bound copy unit at site $i+1$ is capable of receiving a backbone bond from site $i$, then a backbone bond may form. 
\item Depolymerisation, the breaking of backbone bonds, can only occur between template-bound copy units which share a backbone bond, a consequence of this being that units in copolymer tails cannot spontaneously depolymerise.
\end{enumerate} 

We assume that polymerisation or depolymerisation of molecules not attached to the template in the baths or on the end of free copolymer tails is negligible.

\subsubsection{Concentrations and transition energies}

The concentration of monomer units in the baths are set to be equal, $\M_{i}=\M$, for all monomer types $i$, defined as a dimensionless quantity relative to an arbitrary reference concentration. We are interested in copying systems in which any template can be copied with the same efficiency as any other template from the same set of building blocks. Templates with an over-abundance of a certain monomer type could be accurately copied more easily if the corresponding copy monomer was over-represented in the pool too \cite{Poulton2021}. However, a biased monomer pool would, on average, provide no advantage for copying an arbitrary template sequence. 

The monomeric units in the tails of the polymers are constrained to a small volume in closer proximity to the template and neighbouring monomers than the free monomers in the baths. We therefore define an effective concentration $\Meff \gg \M$ in order to parameterize the free energy change of polymerization and the rebinding rate of monomers in the tail to the template.

Bond formation is parameterized through standard free-energy change of reactions at the reference concentration ($\M=1$). In this work we define all free energies as dimensionless quantities relative to $kT=1$. The formation of the backbone bond between monomers in isolation is associated with a standard free-energy change of $\G{BB}+\ln\Meff$; $\G{BB}$ reflects the chemical bond strength and $\ln\Meff$ the loss of entropy associated with binding from solution.  More negative values of $\G{BB}$ favour polymer growth.

Initially in Section \ref{subsec:basic}, we consider a simple model of cooperative copy-template interactions that is depicted in Fig.~\ref{fig:BasicModel}. In this picture, individual monomers bind to the template with a standard free energy change of $\G{spec}$ -- called ``specific" because it depends on whether the copy and template monomers are complementary. The standard free-energy change for binding of a copy polymer of length $l$, forming exactly $l$ bonds with the template is 
\begin{equation}
     \sum_{i=0}^{l-1} \G{spec}^i - (l-1) \ln\left(\Meff/\M\right),
\end{equation} where $\G{spec}^i$ is the specific bond free energy of the $i^{\textrm th}$ copy-template pair. Here the factor $(l-1) \ln\Meff$ captures the cooperative nature of the bond between a long template and a long copy that is responsible for product inhibition. This formulation is an extremely simple model of cooperativity: a copolymer of length $l$ gains $l$ favourable bonds of the same strength but, for $l-1$ of those bonds, the entropic cost is reduced enormously relative to binding from solution.  In Section \ref{subsec:generic} we extend the model to allow for less cooperative behaviour, and in \ref{subsec:weakend} also allow sites at the end of the template to bond more weakly with the copy. 

In the majority of this work we consider a uniform template where there is one type of copy and template monomer, and hence the specific bond strength, $\G{spec}$ takes one value. In the final section of this paper, Section~\ref{subsec:accuracy}, we consider a model with a binary copy monomer alphabet with $\alpha_{C} = 2$. 

\subsubsection{Parameterisation of transition rates}
\label{sec:modelparam}

We assume that each transition described in Section \ref{sec:transitions} and depicted in Fig.~\ref{fig:BasicModel} is well described as an instantaneous process with an average rate. Since the thermodynamics of copy production is an important feature underlying its physics \cite{Ouldridge2017}, we ensure that the thermodynamics of the system is self-consistent by applying the principle of local detailed balance to set the transition rates \cite{Ouldridge2018}. The principle of local detailed balance states that the ratio of forwards to backwards rates between any pair of states is constrained by the chemical free-energy change associated with the forwards transition $\Delta G$, 
\begin{equation}
	\frac{R_{\textrm{forward}}}{R_{\textrm{reverse}}} = e^{-\Delta G }.
\end{equation}
We use this equation to parameterise the transition rates between states in our system, as shown in Fig.~\ref{fig:BasicModel}. 

Monomer binding and unbinding is associated with a free-energy change of $\Delta G = \G{spec} - \ln \M $, since one specific bond is formed in the process of monomer binding. The ratio of the monomer binding rate $R_{\textrm{bind}}$ to the monomer unbinding rate $R_{\textrm{unbind}}$ is  given by 
\begin{equation}
	\frac{R_{\textrm{bind}}}{R_{\textrm{unbind}}} =  \M e^{- \G{spec}}.
\end{equation}

We begin by assuming that monomers bind to the template sites with mass action kinetics, giving $R_{\textrm{bind}} \propto \M$, and that the monomers unbind from the template with a rate that is exponentially dependent on the strength of the specific bond $\G{spec}$, and hence we arrive at 
\begin{equation}
	R_{\textrm{bind}} = k_{0} \, \M \text{ and } R_{\textrm{unbind}} = k_{0} \, e^{\G{spec}} .
\end{equation}

In Section \ref{subsec:accuracy}, we consider the competition between two types of monomer, and compare the effects of two different ways of parameterising the unbinding and binding rates, in which either unbinding rate or the binding rate is exponentially dependent on $\G{spec}$. Prior to Section \ref{subsec:accuracy}, we maintain the parameterisation given above for simplicity.

The free-energy change associated with copy units in the tails at either end of polymers binding to a template site is $\Delta G = \G{spec} -\ln \Meff $. Therefore, the rates of tail binding and unbinding are constrained by
\begin{equation}
	\frac{R_{\textrm{tail bind}}}{R_{\textrm{tail unbind}}} = \Meff e^{ - \G{spec}}.
\end{equation}
Following a similar argument to that for  monomers binding from solution, we set the rate that a unit in a copolymer tail rebinds to an available site on the template to be
\begin{equation}
	R_{\textrm{tail bind}} = k_{0} \, \Meff.
\end{equation}
The rate at which polymer units unbind from a template site to become part of a tail is then identical to the rate at which a monomer in the same position would detach,
\begin{equation}
	R_{\textrm{tail unbind}} = k_{0} \, e^{\G{spec}}.
\end{equation}

In this model, the final detachment and release of a copolymer is irreversible. Release occurs when the last copy-template bond is broken, and therefore we set the rate of release to equal the rate at which monomers unbind from the template, giving
\begin{equation}
	R_{\textrm{release}} = k_{0} \, e^{\G{spec}}.
\end{equation}

Polymerisation between units attached to the template at site $i$ and $i+1$ is associated with a free energy change of $\G{BB}$. Therefore
\begin{equation}
	\frac{R_{\textrm{polym}}}{R_{\textrm{depolym}}} = e^{ -\G{BB}}.
	\label{eqn:gpol}
\end{equation}
We choose $R_{\textrm{polym}} = k$, and therefore $R_{\textrm{depolym}} = k e^{\G{BB}}$. Under this parameterisation, increasing the backbone bond strength increases the average lifetime of a backbone. 

With these assignations for the free-energy change of each reaction, the overall free-energy change of incorporating a single monomer into a polymer in solution is
\begin{equation}
	\G{pol}= \G{BB} - \ln \frac{\M}{\Meff},
\end{equation}
as required.

\begin{figure*}[t]
	\centering
	\includegraphics[width=\textwidth]{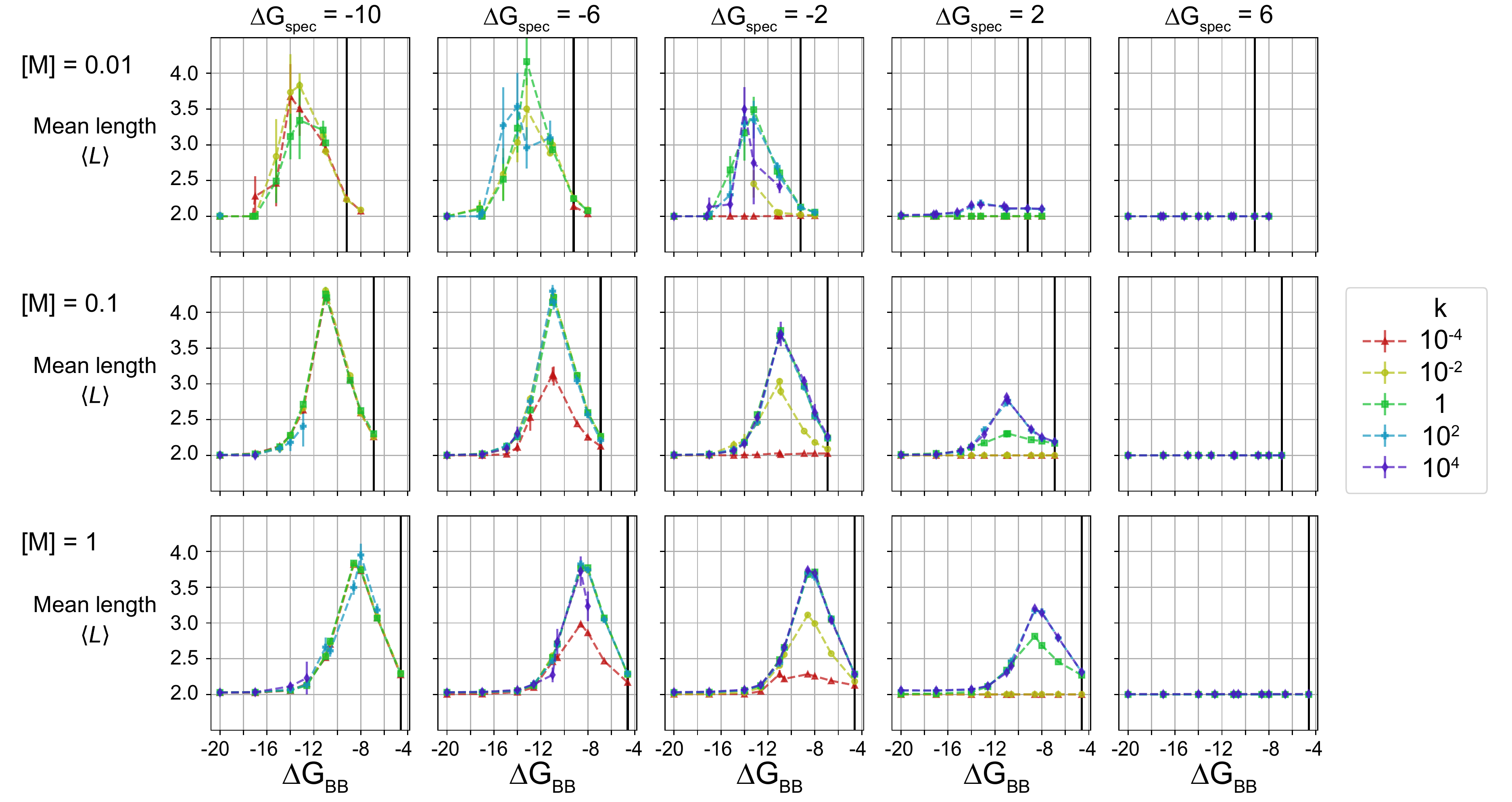}
	\caption{For the basic model of Fig.~\ref{fig:BasicModel}, a wide search over multiple axes of the parameter space shows that the mean polymer length is always low. Each graph shows an average of the mean polymer length produced by the system, $\avg{L}$, with SEM error bars, on a template of length $L_T = 30$, against $ \G{BB}$, with $-\G{BB}$ being the driving force behind backbone bond formation. The monomer concentration $\M$ is varied between rows, and the specific bond strength, $\G{spec}$, between the columns, while the polymerisation rate, $k$, is varied and represented with different colours and marker shapes. Error-bars are smaller than line width where not visible. A black line gives the value of $\G{BB}$ at which $\G{pol} = 0$. In all cases, $\langle L \rangle \ll L_T$, although a moderate peak in $\langle L \rangle$ is observed at moderate values of $\G{BB}$.}
	\label{fig:SimpleModelSweep}
\end{figure*}

\subsection{Methods}
\subsubsection{Gillespie simulation}

We simulate the dynamics of this system using Gillespie's kinetic Monte Carlo simulation method \cite{Gillespie1976} with two levels of stochastic sampling. The first level of sampling selects the template site to update, with a probability proportional to the sum of transition rates at that site, and the second selects the transition rule to apply at that site with a probability proportional to its individual transition rate. This two-stage sampling method is advantageous because the valid transitions of the system only change the rates for subsequent transitions in a localised region of the system. For instance, a monomer binding the template at site 3 will not change the conformation of the system at site 10 or 100, and hence doesn't change the valid transitions around site 10 or 100. Given the locality of the model we have developed, we need only need to recalculate the valid transitions and sum over the valid rates at the few sites either side of the site which was last updated at each step. 

\subsubsection{Parameters, initial conditions and stopping criteria}

We use dimensionless units by setting $k_{B} T = 1$ and $k_{0} = 1$. We also use $\Meff = 100$ throughout. We vary the free parameters $k$, $\M$, $\G{spec}$, $\G{BB}$, and $\G{gen}$ and $\G{end}$ that will be introduced later. Unless stated otherwise, the initial condition of the system in each simulation is an empty template. We simulate systems with templates of length $L_{T} = 10, 30$ and $100$. We calculate statistics of the copying process, such as the mean polymer length and, later when considering systems with two distinct monomer pools, the error rate (defined as the number of mistakes contained within a polymer divided by its length), over the first fixed number of polymers (polymers with length $\geq 2$) that form upon and are released from the template. We take averages of quantities such as the mean polymer length by running multiple simulations with the same parameters, but different random number seeds. Where the number of polymers produced by the system (the sample size) generated within the allotted compute-time for each simulation is less than 10, the data are excluded as they are unrepresentative of steady state. In each case we run between $O(10)$ independent repeats for each set of parameter values, with each independent simulation stopping after producing $O(1000)$ polymers, though this varied from case to case. The code and input files required to reproduce the data presented in this work are freely accessible through the link provided under Data Availability at the end of this paper.

\section{Results}
\subsection{A simple model of copy-template binding and copy-copy polymerization cannot reliably produce long polymers}

\label{subsec:basic}

\begin{figure*}[t]
	\centering
	\includegraphics[width=\textwidth]{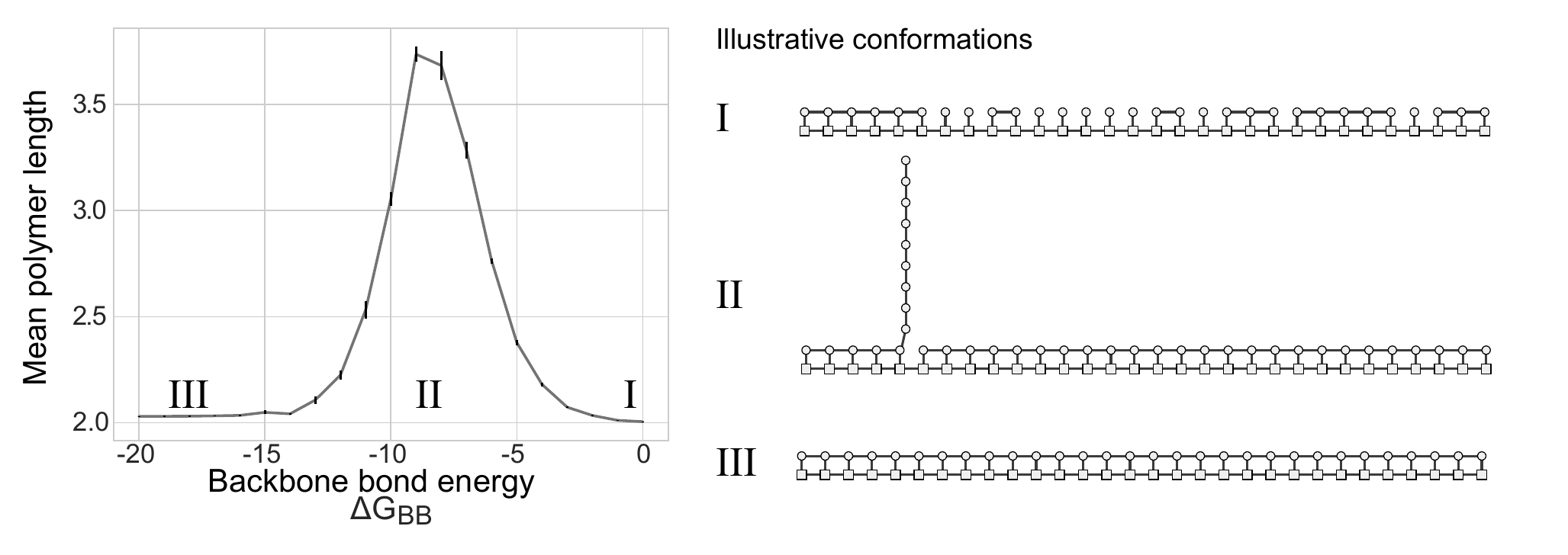}
	\caption{Typical configurations help to explain the observed mean length $\langle L \rangle$ of copy products in the simple model of templating. We plot $\langle L \rangle$ against $\G{BB}$ for a template of length 30,  $ \G{spec} = -4$, $\M=1$ and $k=1$, showing a moderate peak surrounded by regions of $\langle L \rangle=2$. We extract typical configurations at the labelled values of  $\G{BB}$: \textbf{(I)} Weak backbone, many small fragments; \textbf{(II)} Moderate backbone strength, a small number of longer fragments; \textbf{(III)} Strong backbone, a single long polymer covers the whole template.}
	\label{fig:GbbSimple}
\end{figure*}

We begin by asking whether long copy polymers can consistently be produced from a system of monomers governed by the simple model introduced in Section \ref{sec:modelparam}, with transitions parameterised as in Fig.~\ref{fig:BasicModel}. In Fig.~\ref{fig:SimpleModelSweep}, we analyse the polymers produced by the system on a template with length $L_{T}=30$. We show the mean polymer length $\avg{L}$ against the free energy driving backbone formation on the template, $ \G{BB}$ as we vary the polymerisation rate $k = [10^{-4},10^{-2}, 1, 10^{2}, 10^{4}]$, the monomer concentration $\M = [0.01, 0.1 , 1]$, the specific copy-template bond strength $\G{spec} = [-10,  -6,  -2,   2,   6]$, and backbone strength $\G{BB}$ ranging from -4.6 to -20. Five independent simulations were run for each parameter value, with a target of 2000 products. When parameter values where extreme, long embedded Markov processes were observed that were time consuming to simulate. Refer to Section S1 of the SI for a discussion of these parameter values.

As shown in Fig.~\ref{fig:SimpleModelSweep}, the mean length of polymers produced by the system remains low, $2 \lesssim \avg{L} \lesssim 4$, across the parameter space. We now explore particular regions of parameter space to explain why. To help with this analysis, we consider a simple cut through the parameter space, and also probe the actual state of the template during simulations. To do so, we consider the specific case with $ \G{spec} = -4$, $\M=1$ and $k=1$ keeping other parameters the same as in Fig.~\ref{fig:SimpleModelSweep}, and average over 5 independent repeats of 1000 polymers at each data point as we vary the backbone bond strength $\G{BB}$. The results are reported in Fig.~\ref{fig:GbbSimple}, alongside characteristic snapshots from simulations.

\begin{figure}
	\centering
	\includegraphics[width=\columnwidth]{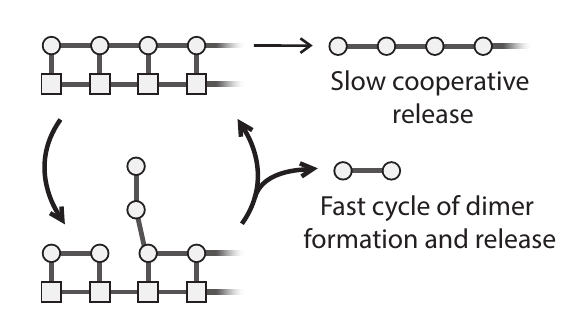}
	\caption{Production of dimers in the limit of strong backbone bonds. Long polymers form on the template when the backbone bond $\G{BB}$ is sufficiently strong. Cooperative binding slows the complete release of the long polymer, while dimers are formed and released under its fraying ends. }
	\label{fig:DimerCycle}
\end{figure} 
Parameter values where a template-bound polymer is unstable with respect to either the dissociation into monomers free in solution ($\G{BB} + \G{spec} - \ln \M >0$), or unconnected template-bound monomers ($\G{BB} >0$) don't feature heavily in Fig.~\ref{fig:SimpleModelSweep} and Fig.~\ref{fig:GbbSimple}, but are consistent with $\avg{L}\rightarrow 2$ as $\G{BB}$ gets more positive as the only ``polymers" produced are transiently bound dimers. 

As $\G{BB}$ becomes less positive/more negative, a significant number of longer polymers appear on the template. An example configuration is shown in Fig.~\ref{fig:GbbSimple}\,I. However, $\avg{L} \approx 2$ remains true, because, although longer copies are present on the template in significant numbers, they tend to detach more slowly than dimers due to the cooperative bond with the template. Therefore dimers dominate the distribution of products. 

At extremely negative values of $\G{BB}$, configurations such as Fig.~\ref{fig:GbbSimple}\, \textbf{III} are obtained. In this limit, a single long polymer covers the whole template, This configuration is extremely stable due to the difficulty in breaking the backbone bond, and the cooperativity with which the long copy binds to the template. However, this cooperativity does not prevent the ends of the long polymer from ``fraying" (undergoing transitory detachment from the template) which allows dimers to form under the raised tails; these dimers are subsequently released into solution must faster than the longer polymer, and are thus registered as the only product (see Fig.~\ref{fig:DimerCycle}).

For some sets of parameters, a moderate peak in $\langle L \rangle$ is observed between these limits. In this case, although $\G{BB}$ is negative, backbone bonds within the copies are broken at an appreciable rate, resulting in relatively long fragments on the template. In some cases, via a mechanism of fraying and polymerization at the junction between fragments, one fragment can displace the other from the template in a step-by-step fashion. An intermediate state in this process is shown in Fig.~\ref{fig:GbbSimple}\,\textbf{II}. We observe that this peak occurs at slightly negative values of $\G{pol}$, which can be rationalised by noting that adding a single monomer to the tail of template-attached polymer is associated with a free-energy change of $\G{pol}$, and so states with long tails are thermodynamically unfavourable for $\G{pol}>0$.

Although this mechanism can provide some increase in  $\langle L \rangle$, a longer polymer is more likely to push a shorter one off the template, and no complete polymers are produced. Importantly, therefore, even the systems that give $\langle L \rangle \neq 2$ still produce outputs that are much smaller than the template, and do not have a well-defined length. Either template-bound polymer formation is too unfavourable, or cooperative template binding prevents the release of long stable polymers (product inhibition). Putative displacement mechanisms can be initiated at any location and produce fragmented polymers with a bias towards shorter products. These results suggest that, for systems that are well-described by the basic model of Section \ref{sec:modelparam}, the basic templating mechanism is incapable of generating and, crucially, releasing long polymers reliably. Indeed, in Ref.~\cite{Rosenberger2021}, in a bottom-up, coarse-grained model of a replicating system that obeyed similar dynamics to that described by the model in Fig.~\ref{fig:BasicModel}, the authors observed the formation of only long polymers that remained bound in complexes under isothermal conditions.  

In the next section, we extend the model to allow for a more complex polymerisation mechanism. This scheme can be viewed simultaneously as a way to avoid product inhibition, and a way to favour effective displacement mechanisms, by disrupting the cooperativity of copy-template interactions in a directional manner.

\subsection{A polymerisation mechanism that disrupts copy-template binding allows for long copies}
\label{subsec:generic}

\begin{figure*}[t]
	\centering
	\includegraphics[width = \textwidth]{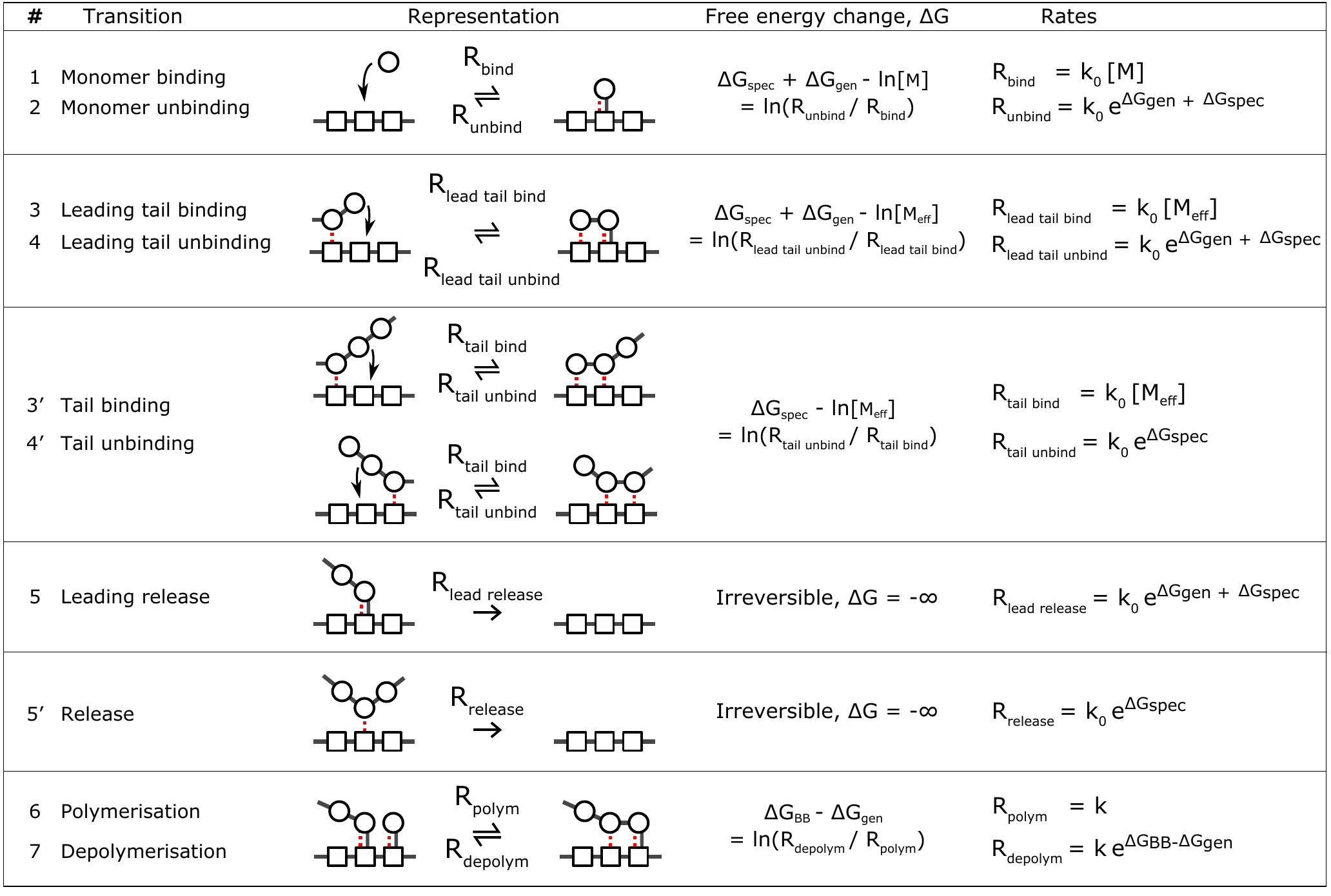}
	\caption{The generic bond model is an extension of the basic model presented in Fig.~\ref{fig:BasicModel}. Here there are two kinds of copy-template bond, the specific bond (dashed red line) that is potentially sensitive to the match between copy and template units, and the generic bond (solid black line), that is not sensitive to the match. The generic bond is formed during monomer binding and breaks upon polymerisation, and {\it vice versa} for the reverse reactions. Units at the leading edge of copy polymers \markup{(at the right-hand edge)  bind to the template more strongly than other monomers in copies due to} the modified polymerisation mechanism. The generic bond energy, $\G{gen}$, like the specific bond energy, $\G{spec}$, doesn't contribute to the overall free energy change of polymer extension $\G{pol}$. }
	\label{fig:GenericModel}
\end{figure*}

In the context of simple catalysts, Ref.~\cite{Deshpande2020} argued that product inhibition can be reduced if part of the free energy of product formation is diverted into destabilizing the interaction of catalyst and product. With that idea in mind, we develop a model in which backbone bonds in the copy can only be formed at the expense of breaking a bond between copy and template. The scheme, illustrated in Fig.~\ref{fig:GenericModel}, could describe a range of chemical systems with competitive bond formation, but in particular resembles the handhold-mediated strand displacement mechanism introduced by Cabello-Garcia \etal \cite{Cabello-Garcia2021}.

Formally, we implement this mechanism by splitting the copy-template bond into two parts, one of which must be broken for a backbone bond to form. The free-energy change of monomers binding to the template from solution is now given by $\Delta G = \G{spec}+\G{gen}- \ln\M$, where the generic bond $\G{gen}$ is not dependent on the match between copy and template. We assume that the polymers have a directional asymmetry (as is typical in macromolecular polymers like DNA and RNA \cite{MolBiolCell}), and represent this in diagrams such as Fig.~\ref{fig:GenericModel} by drawing the specific bond (dashed) on the left and the generic bond (solid) on the right.  \markup{This asymmetry will be retained throughout the manuscript.} 

As shown in reactions 6 and 7 of Fig.~\ref{fig:GenericModel}, the generic bond in the leftmost monomer breaks when polymerisation occurs between neighbouring monomers, leaving the leftmost monomer less tightly bound to the template. \markup{In visual representations, polymerisation will always disrupt the bond of the left or `lagging' monomer; the  generic bond of the right or `leading' monomer is unaffected.} The total free-energy change of the template-attached polymerisation step is then $\Delta G = \G{BB} - \G{gen}$. $\G{gen}$, like $\G{spec}$, does not contribute to the the overall free energy change of extending a polymer tail (or a polymer in solution), $\G{pol} = \G{BB} - \ln \M / \Meff$. \markup{We assume that the disruption of the generic bond during polymerisation happens at the same time as the polymerisation reaction; {\it i.e.,} the bond is transferred from template to polymer via an ``attack" or an ``invasion" mechanism, as occurs in Ref.~\cite{Cabello-Garcia2021}.}

Having modified the thermodynamics of the model, we modify the kinetics as follows. As shown in Fig.~\ref{fig:GenericModel}, generic bonds are only formed during the binding of monomers from solution, when the leading monomer in a copy polymer rebinds to the template from a tail state, or when the backbone between two template-attached monomers is disrupted. For simplicity, we assume that all unbinding reactions that break a generic bond have rates that scale as $R_{\textrm{unbind}} \propto e^{\G{gen}}$ (so $\G{gen}$ does not appear in the binding rates), and that $R_{\textrm{depolym}} = k e^{\G{BB} - \G{gen}}$, $R_{\textrm{polym}} = k$.

\begin{figure*}[t]
	\centering
	\includegraphics[width=\textwidth]{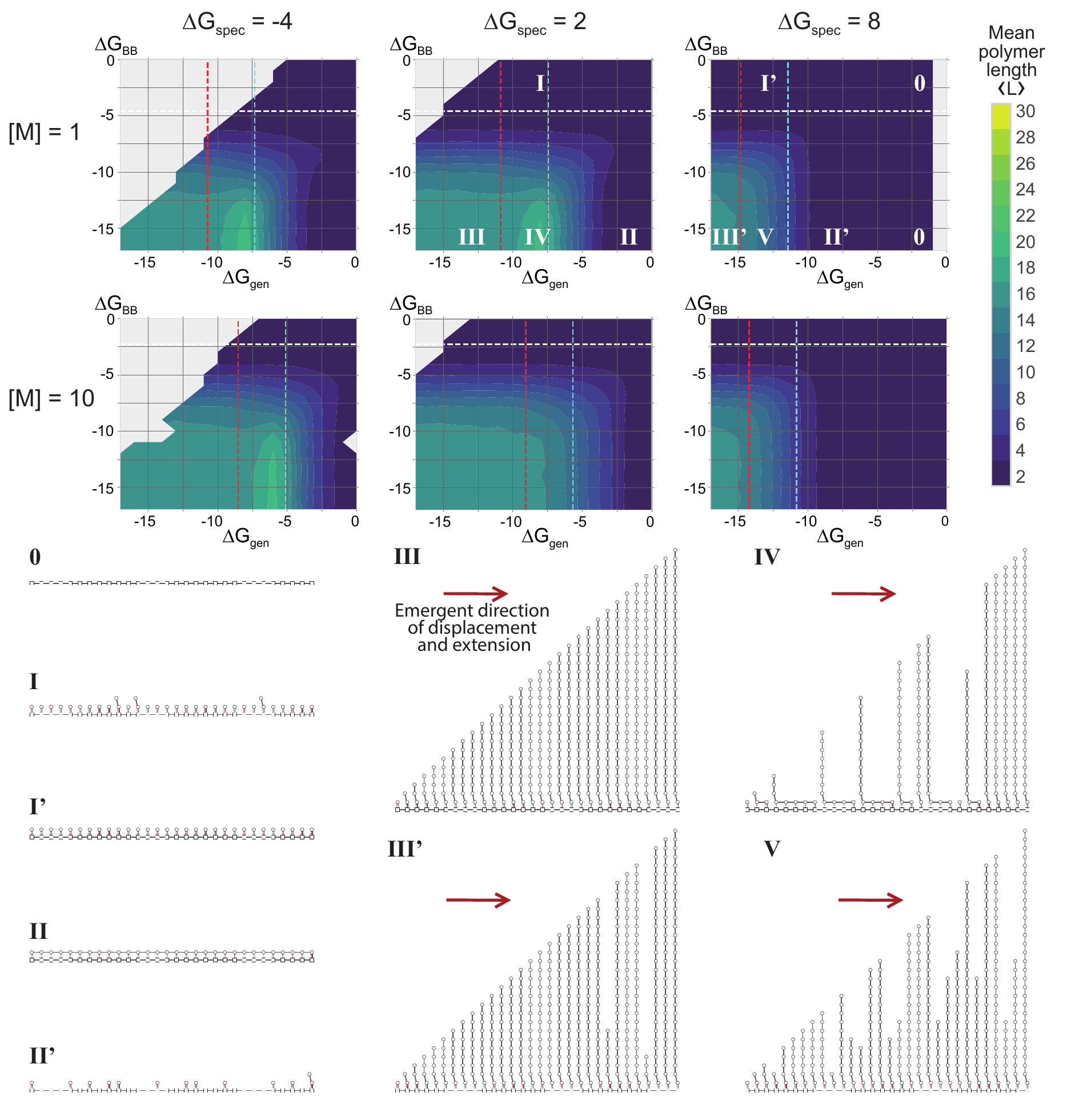}
	\caption{Coupling polymerisation to disrupting copy-template bonds can generate copies with average length $\avg{L}\sim L_{T}/2$. Above, we show surface plots of the mean polymer length, $\avg{L}$, against the backbone bond strength $\G{BB}$ and the generic bond strength $\G{gen}$ for monomer concentrations $\M = {1,10}$ and $\G{spec} = {-4,2,8}$ on a template of length $L_T = 30$. Below, we provide snapshots of the system for the indicated points within the parameter space as the simulation time passed $t = 200000$. White dashed line constraint: $\G{BB} = \ln (\M/\Meff)$. Above the white line, the template is either occupied by monomers and dimers (\textbf{I} and \textbf{I'}) or is empty \textbf{0} because the backbone bond is too weak. The system is only capable of producing dimers $\avg{L} \approx 2$. Dashed cyan line constraint: $\G{gen}  < -\G{spec} + \ln \frac{2 k_0}{L_{T} \ts{disp}}$. To the left of this line, the generic bond is strong enough that a lagging polymer tends to step forward faster than it detaches. To the right of the cyan constraint, we see either an empty template at \textbf{0} (since everything detaches rapidly) or, when the specific bonds are stronger, a template with a fully bound copy, since invading copies cannot force it off the template (configuration \textbf{II}). Here the system still produces dimers $\avg{L} \approx 2$. Dashed red line constraint: $\G{gen} < -\G{spec} + \ln \frac{2 k_0}{L_{T}^2 \ts{disp}}$. To the left of this constraint, lagging polymers step forwards fast enough to create a dense, orderly polymer brush, as in \textbf{III} and \textbf{III'}. A uniform product length distribution is observed with a mean length $\avg{L} = L_T/2$. In between the red and cyan constraints, at \textbf{IV}, slower advancement of the lagging polymers leads either to a less dense brush with some tails occupying the template (IV), or the spontaneous appearance of gaps that encourage the initiation of shorter polymers within the brush (V). In configurations such as IV, $\avg{L} \gtrsim L_T/2$ is observed, whereas for parameters that lead to configurations like V, the mean length is slightly lower $\avg{L} < L_T/2$.}
	\label{fig:Generic}
\end{figure*}
\begin{figure*}[t]
	\centering
	\includegraphics[width=0.84\textwidth]{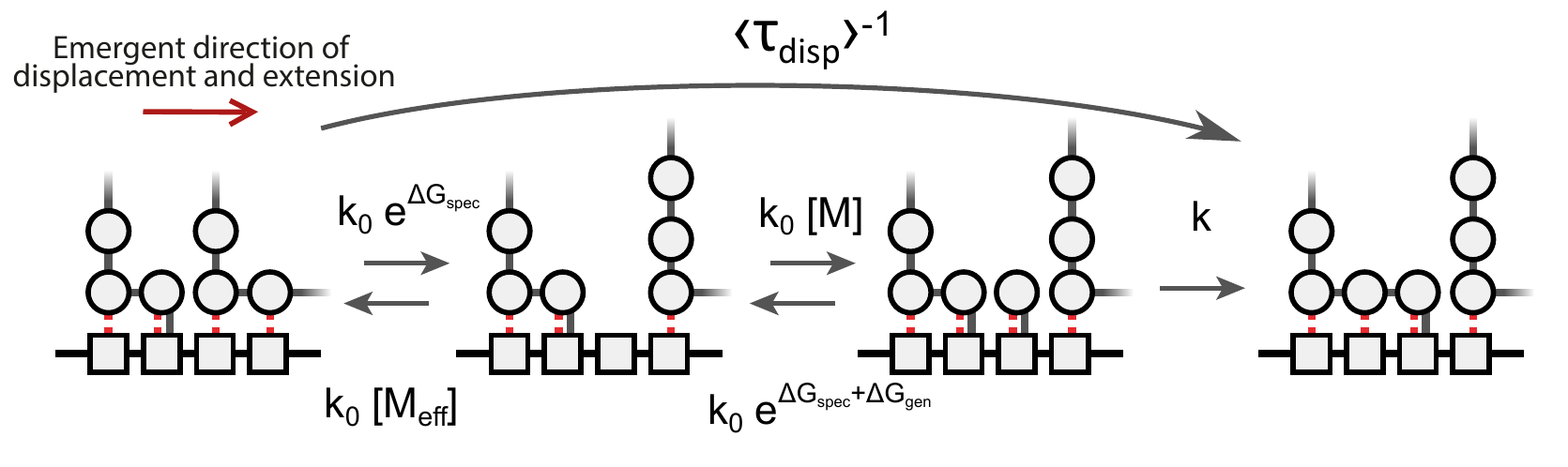}
	\caption{Model to estimate the time, $\ts{disp}$, taken for a lagging polymer to extend by one unit, displacing the tail of the polymer ahead.}
	\label{fig:GenStep}
\end{figure*}

In Fig.~\ref{fig:Generic}, we present multiple contour plots of the averaged mean length of polymers produced via a template of length $L_T = 30$ as the backbone bond strength $\G{BB}$ and the generic bond strength $\G{gen}$ take values in the range $[0,-1,...,-17]$. For each point sampled, we averaged the mean length of 1000 polymers each produced by 5 independent simulations. Data are excluded for simulations that didn't produce this 1000 polymers within the allowed window. We set $k = 1$, and consider $\M = [1,10]$, $\G{spec} = [-4,2,8]$. Additionally, we provide snapshots showing the wide variety of conformations reached by the system at a selected points in the parameter space when the simulation time passed $200,000$ units. 

In Fig.~\ref{fig:Generic}, we observe a high mean length of around $L_{T}/2 = 15$ over large regions of parameter space. This behaviour corresponds to a fully-occupied template, with a ``brush" of polymers attached to it (regions exemplified by \textbf{III}, \textbf{III'}, \textbf{IV} and \textbf{V}). In all of these regimes, the coupling of polymerisation to the disruption of generic bonds with the template has two important effects. Firstly, provided the disrupted bond is strong enough, the mechanism can overcome the cooperative effect that leads to product inhibition by a single long polymer. The leading monomer of a copy binds strongly, but the lagging monomers bind much more weakly. Undesirable confirmations with a single long polymer bound to the template are recovered (see configuration \textbf{II} in \ref{fig:Generic}) when $\G{gen}$ is not large and negative, but the specific bond is strong. Secondly, the asymmetry introduced by the generic bond  solves the problems associated with the unbiased strand displacement mechanisms mention in Section \ref{subsec:basic}. The leading edge of a lagging polymer out-competes the tail of the polymer ahead by binding to the template with greater strength at their junction, allowing the system to act as a ratchet in which biased strand displacement causes shorter, lagging polymers to push longer polymers ahead off the template. \markup{The emergent direction of displacement and extension is indicated in Fig.~\ref{fig:Generic} by a red arrow}. As shown in configuration \textbf{III} of \ref{fig:Generic}, the result is a brush of polymers connected to the template by only one strong bond across the length of the template. The random detachment of polymers from this brush produces a distribution with an average polymer length $\langle L \rangle \sim L_{T}/2 =\sim 15$.  

Some simple physical reasoning can be used to explain the regions of parameter space in which various behaviours are observed. For long polymer copies to form easily, incorporating monomers into copy polymers  must be thermodynamically favourable \cite{Poulton2021}. Thus $\G{pol} = \G{BB} - \ln (\M/\Meff)<0$, and we require a sufficiently strong backbone: 
\begin{equation}
	\label{eqn:whiteconstr}
	\G{BB} \ll \ln (\M/\Meff).
\end{equation}
The line representing this constraint is shown in white in Fig.~\ref{fig:Generic}. Above this line, the system exhibits either an empty template as in configuration \textbf{0} if the copy-template interaction if weak ($\G{spec}+\G{gen} > \ln  \Meff $); or a template largely covered in monomers, as in \textbf{I} and \textbf{I'}, if the copy-template interaction is strong ($\G{spec}+\G{gen} < \ln  \Meff $). When the constraint of Eq.~\ref{eqn:whiteconstr} is not satisfied, the system is only capable of producing short polymers, much for the same reasons we presented in Section \ref{subsec:basic} and Fig.~\ref{fig:GbbSimple}, as long polymers are unstable structures. 

Given a sufficiently stable backbone, long polymers can form. However, the strength of the generic bond $\G{gen}$ can drastically change the conformations that these long polymers settle into, and thereby change the typical length of polymers that are produced by the system. When the generic bond is weak, as in  \textbf{II} where $\G{gen} = -2$, we tend to observe single, long polymers stuck on template, and dimers are created and released under the fraying end of the polymer as in \ref{subsec:basic}. Note that when $\G{gen} = 0$, the system reverts back to the model presented in Section \ref{subsec:basic}. When the generic bond is strong and takes very negative values, as in \textbf{III} and \textbf{III'}, we observe conformations where the polymer tails form a dense brush, and the mean length is roughly half the template length $\avg{L} = L_{T}/2$. We also observe regions with configurations like \textbf{IV} and \textbf{V}, in which the copy polymers form a less dense, less regular brush.

The key criterion for whether a brush of long polymers forms relates to two timescales. The first is the time taken for a  polymer to polymerise forward into a space that is occupied by the lagging edge of the polymer in front. We call this time $\ts{disp}$, since the lagging tail of the polymer in front is effectively displaced (although in many cases its binding is weak to begin with). The second is the timescale on which the leading edge of a polymer detaches, $\ts{unbind}$. We can estimate $\ts{disp}$ with the simple discrete-state, continuous-time model in Fig.~\ref{fig:GenStep}. We depict the four states involved in a single forward step alongside their transition rates. Treating this subprocess in isolation, we obtain a rate matrix $\textbf{K}$ with entries:
\begin{equation*}
	\textbf{K} = 
	\begin{pmatrix}
		-k_0 e^{\G{spec}}   &  k_0 \Meff            & 0 & 0 \\
		k_0 e^{\G{spec}}    & - k_0 \Meff - k_0 \M & k_0 e^{\G{gen} + \G{spec}} & 0 \\
		0                          & k_0 \M                   & -k_0 e^{\G{gen} + \G{spec}}-k & 0  \\
		0                          & 0                         & k  & 0 
	\end{pmatrix}
\end{equation*}

The mean first-passage time $\ts{disp}$ from state 0 to state 3 is given by the $(3,0)$ element of the Drazin inverse of $\textbf{K}$, which takes a value

\begin{align}
\label{eqn:tdisp}
    \ts{disp} = &( k (1 + (\M + \Meff)e^{-\G{spec}})   \\
     + &k_0 (e^{\G{gen}+\G{spec}} + \M + e^{\G{gen}} \Meff))/(k k_0 \M), \nonumber
\end{align}

$\ts{unbind}$ is simply given by the inverse of the monomer unbinding rate, $\ts{unbind} = (R_{\textrm{unbind}})^{-1} = \frac{1}{k_0} e^{-\G{gen} - \G{spec}}$. To produce copies of length $\sim L_{T}/2$, it is necessary that $\ts{disp} L_{T}/2 < \ts{unbind}$, otherwise the growing polymer will detach too rapidly. By rearranging this inequality, we arrive at
\begin{equation}
\label{eqn:constrdispl}
	\G{gen}  < -\G{spec} + \ln \frac{2 k_0}{L_{T} \ts{disp}}
\end{equation}
which can be implicitly solved for $\G{gen}$. This constraint is represented by the cyan dashed line in Fig.~\ref{fig:Generic}. When this constraint isn't satisfied, newly-initiated polymers tend to detach before they extend. When the total copy-template affinity $\G{spec}+\G{gen}$ is strong, as at \textbf{II}, we observe a single, long polymers stuck to the template, and when when the affinity is weak, at \textbf{0}, we observe empty templates. Polymer brush configurations \textbf{III}, \textbf{III'}, \textbf{IV} and \textbf{V}, which all produce a relatively long distributions, all lie to the right of this constraint.

For a dense, ordered brush conformation as in \textbf{III} and \textbf{III'} to be stable, $\sim L_T/2$ sequential extension and displacement steps must occur on a timescale shorter than that at which any of the $L_T$ polymers fall off, enabling the system to heal the brush after any detachment events. Therefore, a dense, brush-like conformation will be reached when $\ts{\rm disp}  L_{T}/2 < \ts{\rm unbind}  /L_T$. Rearranging, we obtain
\begin{equation}
\label{eqn:constrdense}
	\G{gen} < -\G{spec} + \ln \frac{2 k_0}{L_{T}^2 \ts{disp}},
\end{equation}
represented by the red dashed line in Fig.~\ref{fig:Generic}. In a maximally-displaced, brush-like conformation, all polymers are only attached by their leading edge. Hence detachment can occur equally from any point and the product length distribution is uniform, with a mean length very close to $\avg{L} = L_T/2$.

When $\G{gen}$ lies between the constraints Eqs.~\ref{eqn:constrdispl} and \ref{eqn:constrdense}, the generic bond is strong enough to enable long polymers to form but insufficient to reach a maximally-displaced, brush-like conformation. If the specific bond strength is strong enough that lagging tails tend to bind back to the template, and are only removed via displacement, we see conformations similar to \textbf{IV}. Here, we see gaps between the leading edges of polymers, with weakly-bound tails occupying the intervening sites.  Some shorter polymers occupy large swathes of the template, while some longer polymers have been pushed to the end of the template where they can advance no further. Here we see a slight inversion of the cooperative effect; longer polymers which bunch at the end of the template tend to have fewer template connections than shorter polymers, which skews the product length distribution slightly toward long polymers giving $\avg{L} > L_T/2$.

\begin{figure*}[t]
	\centering
	\includegraphics[width = \textwidth]{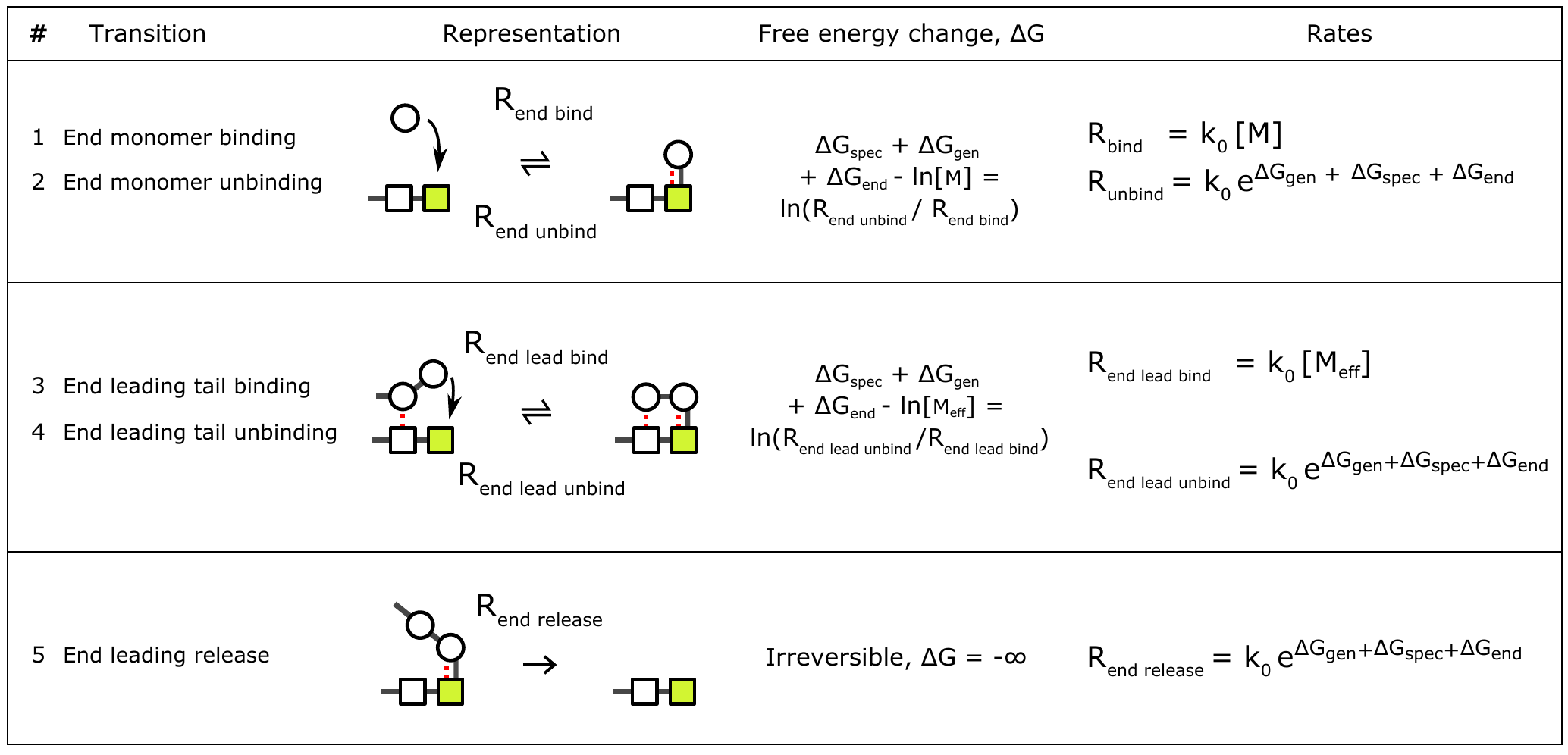}
	\caption{The dynamics of the system at the weakened end (green square) of the template. We extend the generic bond model presented in Fig.~\ref{fig:GenericModel} and modify the copy-template interaction strength at the end of the template with an energetic term $\G{end}$. When $\G{end}>0$, the bond made between copy units and the final template site is weakened. \markup{As before, the right hand edge of the copy polymer is the leading edge.} }
	\label{fig:EndModel}
\end{figure*}

This effect is not observed when the specific bond is weak enough that polymer tails can spontaneously detach from the template without the need for displacement, as seen at \textbf{V} where $\G{spec} = 8 > \ln{\Meff}$. Here gaps appear quicker than they can be filled by extension of polymers already on the template, and  monomers that bind within these gaps can sometimes form dimers. The tails of the dimers may spontaneously detach from the template, preventing the incorporation of the dimer into a the polymer behind, and initiating a new copy from the centre of the template. As a consequence, the ordered polymer brush conformation of \textbf{III} or \textbf{III'} is disrupted by shorter polymers that have incorrectly initiated in the middle of the template. Consequently, the system produces polymers with a mean length $\avg{L} < L_T/2$.

We have not observed any parameter values in which the template sites are typically unoccupied but in which long copies are frequently produced. To get long polymers, monomer binding in isolation must be stable ($R_{\rm bind}/R_{\rm unbind} >1$) since we have deliberately disrupted cooperative binding. Therefore the only way that the template sites can typically be available is if both (a) lagging tails tend to unbind, and (b) if stepping forward is too fast for monomers to fill in behind. But a monomer filling in happens at least as quickly as a polymer can step forwards, by definition, and much faster if the polymer runs into a traffic jam of other polymers. So empty sites tend to be a rare commodity unless binding to the template is just pathologically unstable.

We have found that applying the two constraints of Eqs.~\ref{eqn:whiteconstr} and \ref{eqn:constrdispl} is generally sufficient to identify the region in which long copies are produced (Fig.~\ref{fig:Generic}). We note that in the upper left triangle of Fig.~\ref{fig:Generic}, where $G_{BB} \gg G_{gen}$, the system failed to produce 1000 polymers during the allotted run-time and the data were excluded. When $G_{BB} \gg G_{gen}$, the depolymerisation step in the model becomes very fast causing repeated polymerisation and depolymerisation events. In this region, the embedded process of polymer growth becomes very long and requires ever longer run-times to simulate. In the under-sampled regions, we have no evidence to suggest that the polymer length distributions produced by the system deviate from the simple arguments presented above. 

By introducing a mechanism that channels the free energy released during polymerisation into destabilising the interactions of polymers tails and the template, we have shown that far longer polymers can be produced in a simple model of templating. However, for true copying, the polymers produced by the system must have identical lengths to the template. In the following section, we bias the production complete polymers simply by weakening the copy-template interaction strength at end of the template.
\vspace{0mm}
\subsection{Weakening the final site on the template biases the production of complete polymers}
\label{subsec:weakend}

\begin{figure*}[t]
	\centering
	\includegraphics[width=0.7\textwidth]{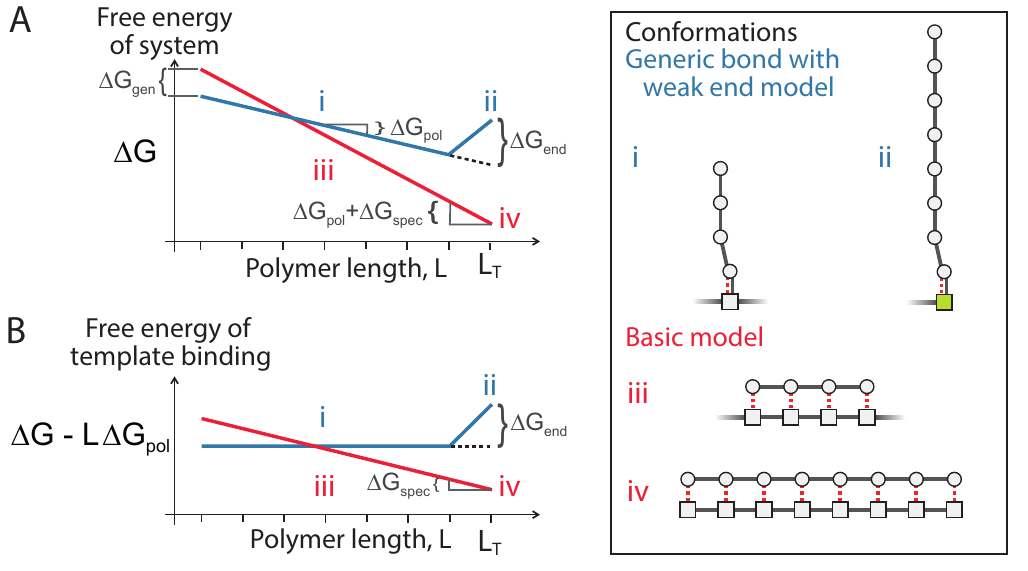}
	\caption{\markup{ \textbf{A.} The free energy of a polymer, $\G{}$, decreases with length, $L$ when $\G{pol}<0$ in all models. \textbf{B.} The component of the free energy due to the copy-template bonds is given by $\G{} - L\, \G{pol} $. The copy-template free energy decreases with length in the basic model (red line and conformations iii and iv) which suppresses the release of long polymers. In the maximally displaced conformations (i and ii) generated under the generic bond model (blue line), all polymers are only connected to the template at their leading edge, and hence the free energy landscape is flat. Weakening the final site on the template by an amount $\G{end}$ destabilises polymers with length $L = L_T$ (ii), leading to their selective release. } }
	\label{fig:FreeEnergy}
\end{figure*}

We now propose a simple adjustment to the model presented in Section~\ref{subsec:generic} that allows for the systematic release of complete polymers from the template when the system is in a brush-like conformation. Here, the copy-template bond at the final template site is weakened with an free-energetic factor $\G{end}>0$. This change is somewhat analogous to a stop codon in translation \cite{Rodnina2018}. Crucially, since the end of the template is a unique site, this change can be made in a way that doesn't interfere with the copy-template interactions at other sites. 

\markup{ The effect of this adjustment is illustrated in the free energy profiles shown in Fig.~\ref{fig:FreeEnergy}. In the previous section we saw that disrupting the copy-template bond upon polymerisation could allow the lagging tail of polymers to detach from the template, meaning that, although polymerisation was thermodynamically "down-hill", the binding strength of the copy to the template did not increase as the copy extends; the template binding free energy landscape was flat. Here, by weakening the copy-template interaction strength at the end of the template, $\G{end}$, we expected to destabilise the binding of completed polymers and promote their release over incomplete polymers. }

The thermodynamics and kinetics of the modified system are identical to the model presented in Section~\ref{subsec:generic} and Fig.~\ref{fig:GenericModel}, apart from at the very end of the template. In Fig.~\ref{fig:EndModel}, we outline the thermodynamic and kinetic changes to the model. The free-energy change of  binding  at the end of the template is now given by $\Delta G = \G{spec}+\G{gen}+\G{end}- \ln\M$, and the free-energy change of polymer tail binding at the end of the template is now given by $\Delta G = \G{spec}+\G{gen}+\G{end}- \ln\Meff$. The rates of monomer unbinding, leading-edge polymer unbinding and polymer termination at the end of the template all take the value $R = k_0 e^{\G{spec}+\G{gen}+\G{end}}$; binding rates are unchanged relative to Section~\ref{subsec:generic}.

\begin{figure*}[t]
	\centering
	\includegraphics[width=\textwidth]{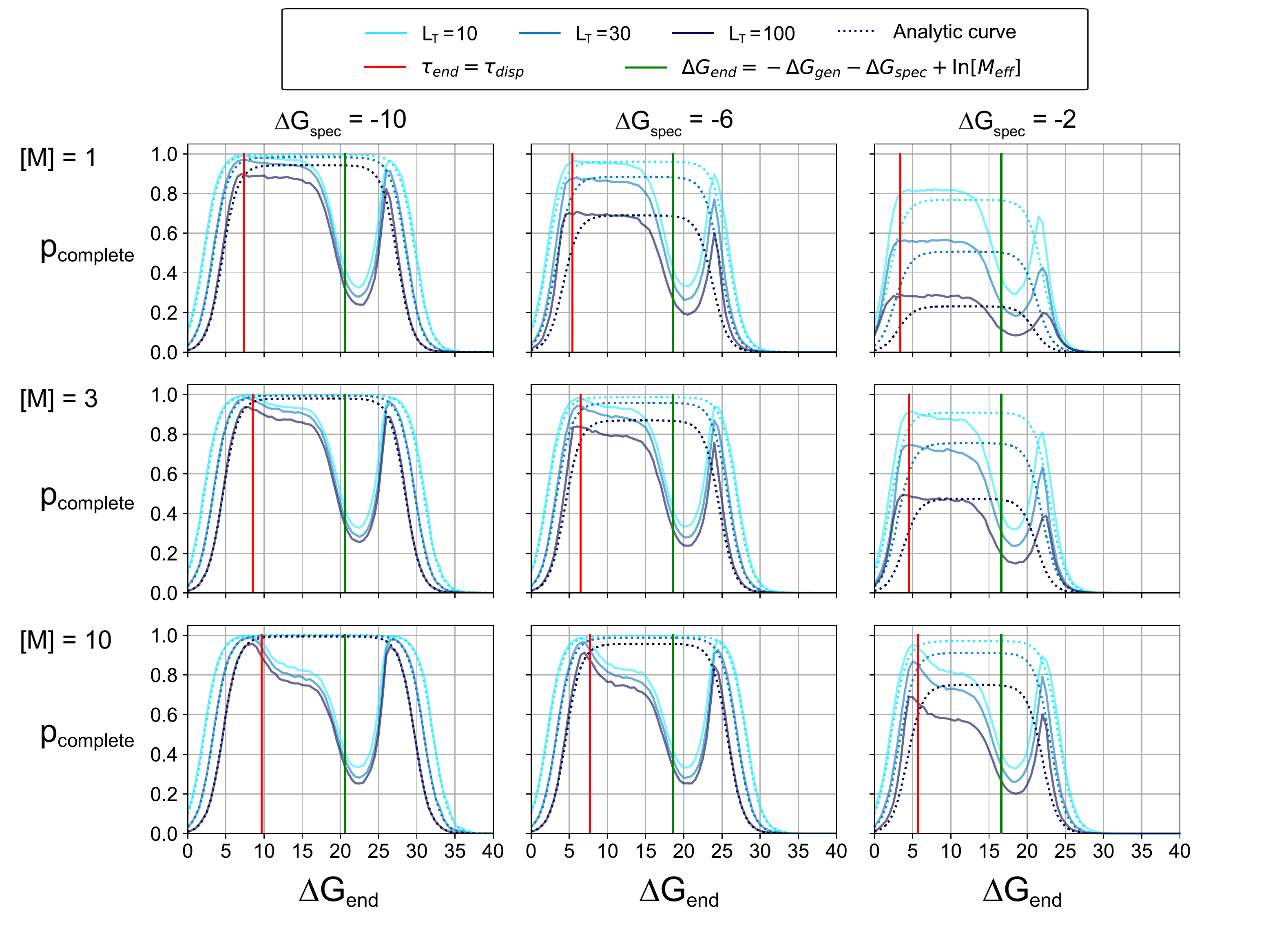}
	\caption{Destabilising the last site on the template with $\G{end}$ increases the probability of producing a complete polymer $\pcomp$ when the configuration of copies forms a dense brush as seen in conformation \textbf{III} in Fig.~\ref{fig:Generic}. We plot the averaged probability of producing a complete polymer, $p_{\rm complete}$, against the destabilising energy penalty, $\G{end}$, for a range of system parameters. We compare the data to a simple analytical model that accounts for the flux of complete polymers and uniformly distributed polymers from the template, and also indicate the points at which $\ts{disp}= \ts{end}$ (red line) and $\G{end} = -\G{gen}-\G{spec} + \ln\Meff$ (green line) to guide interpretation, as explained in the text. }
	\label{fig:Gend}
\end{figure*}

\begin{figure}
	\centering
	\includegraphics[width=0.95\columnwidth]{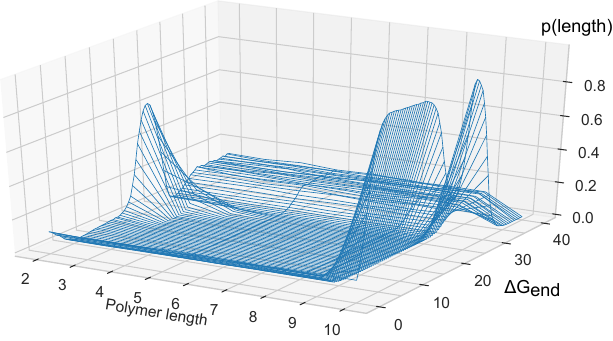}
	\caption{ The full length distributions of the data shown in Fig.~\ref{fig:Gend} for $\M = 1$ and $\G{spec} = -4$ and a template of length $L_T = 10$, averaged over 5 independent simulations at each value of $\G{end}$. At $\G{end} = 0$, we see a approximately uniform product distribution. At low $\G{end}$ we see a sharp increase in the probability of producing complete polymers with $\G{end}$ before a plateau is reached. At higher $\G{end}$, a peak in the production of dimers is observed. As $\G{end}$ is increased further, there is a final peak in the production of full length products, after which the system produces a uniform distribution of products excluding complete polymers. }
	\label{fig:GendDist}
\end{figure}

\begin{figure*}[t]
	\centering
	\includegraphics[width=\textwidth]{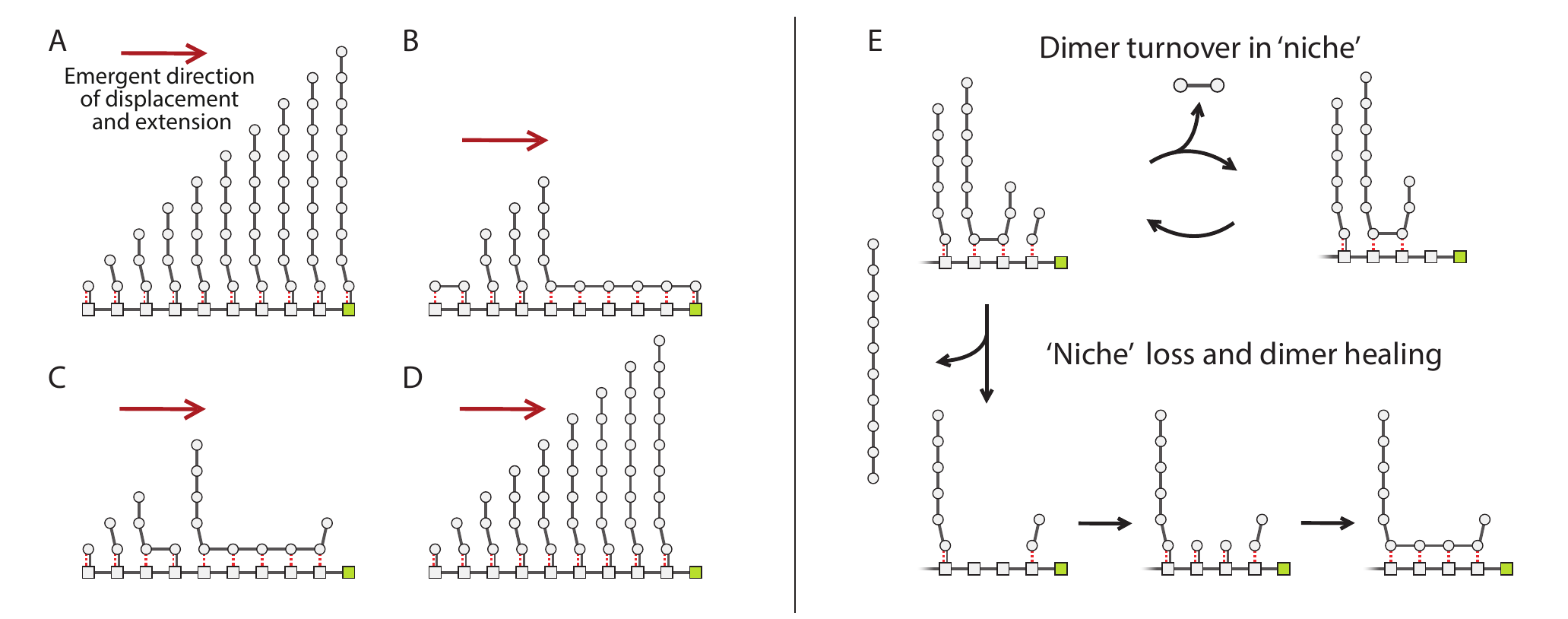}
	\caption{ Weakening the end of the template with $\G{end}$ affects the typical conformations of the system on the templates. We illustrate typical configurations from the simulations reported in Fig.~\ref{fig:GendDist}. \textbf{A} At $\G{end} = 0$, the system reaches a dense brush. \textbf{B} At $\G{end} = 10$, completed polymers are released sooner than displacement can compensate, and so the brush of copies becomes less dense. \textbf{C} At $\G{end} = 22$, the final template site has low occupancy. Here dimers form under the fraying ends of completed polymers. \textbf{D} At $\G{end} = 35$, the occupancy of the final site is so low that incomplete polymers detach sooner than they can complete by polymerising at the end of the template. \textbf{E} Niche formation. The leading edge of completed polymers may fray, creating a protective niche that promotes the cyclical turnover of dimers. If the completed polymer falls off faster than a dimer can form, then the dimer can be incorporated into the preceding polymer.    }
	\label{fig:GendConfDimer}
\end{figure*}

In Fig.~\ref{fig:Gend}, we plot the probability of producing a full-length copy polymer, $\pcomp$ against the destabilising free-energy penalty of the last site, $\G{end}$, at a few values of the generic bond strength $\G{gen}$, the monomer concentration $\M$ and the template length $L_{T}$, as a function of $\G{end}$. We set $k = 1$, $\G{spec} = -4$ , and $\G{BB} = -20$, which puts the system into a dense brush steady state conformation when $\G{end} = 0$ as seen in Section~\ref{subsec:generic}. For each parameter point sampled we ran 5 independent repeated simulations that would produce up to 2000 polymers or truncated after 4hrs run-time. In Fig.~\ref{fig:GendDist}, we show the full product length distribution for $\M = 1$, $\G{spec} = -4$, $L_T = 10$.

Fig.~\ref{fig:Gend} shows that weakening the end of the template can increase the proportion of complete products, $\pcomp$, and that $\pcomp \approx 1$ is possible under certain conditions. When $\G{end} = 0$, the polymer brushes encountered in Section~\ref{subsec:generic} tend towards a jammed state, as depicted in Fig.~\ref{fig:GendConfDimer} \textbf{A}, with detachment events occurring at any point on the template. Weakening the final site on the template creates a release point that allows this polymer `traffic' to flow through, with unbinding predominantly occurring at the final site where complete polymers are attached by a weakened final copy-template bond, \markup{ as seen in the binding free energy landscape in Fig.~\ref{fig:FreeEnergy}.}

To probe this behaviour in more detail, we construct a simple model to explain the proportion of complete polymers in the product pool, $\pcomp$, produced by the system when a dense conformation has been reached. This model considers only the rate of formation and release of complete polymers and the rate of a competing unwanted process of release of a uniform distribution of polymers from the body of the template. In the densely-packed regime, as depicted in Fig.~\ref{fig:GendConfDimer} \textbf{A}, all polymers have one bond with the template. The timescale on which one of $L_T - 1$ incomplete polymers are released from a template of length $L_{T}$ is approximately,
\begin{equation}
	\ts{uniform} = (L_{T} - 1)^{-1} R_{\rm lead\,\,release}^{-1} \approx (L_{T} - 1)^{-1} \, k_0 \, e^{-\G{gen}-\G{spec}}.
\end{equation}
Next we estimate the time it takes the system to complete a growing polymer and release it from the template.To extend the polymer attached to the penultimate site, a free monomer must occupy the final site. The fraction of time that this final site is occupied (as opposed to empty), $g$, can be approximated by the binding equilibrium of the monomer 
\begin{equation}
	g \approx \frac{k_0 \, \M}{k_0 \, \M+  k_0 \, e^{\G{gen}+\G{spec}+\G{end}}}
\end{equation}
Given a polymerisation rate $k$, the average timescale on which a polymer with length $L_{T}-1$ is completed by polymerising with a monomer at the end site on the template is
\begin{equation}
	\ts{form} = 1/ g k \approx \frac{\M+ e^{\G{gen}+\G{spec}+\G{end}}}{\M \, k}
\end{equation}

In this simple model we assume that the time it takes this newly completed polymer to fall off the template is rate limited by either the timescale on which the bond with the end of the template is broken, 
\begin{equation}
	\ts{end} = 1/ k_0 \, e^{\G{gen}+\G{spec}+\G{end}},
\end{equation}
or the time it takes for the polymer unit at the penultimate template site to be displaced by a polymer behind, which is given by $\ts{disp}$ as calculated in Eq.~\ref{eqn:tdisp} in Section \ref{subsec:generic}.

Taken together, the timescale on which complete polymers fall off the template is approximately
\begin{equation}
	\ts{fall} \approx \ts{end} + \ts{disp},
\end{equation}
and the total timescale of producing and releasing a complete polymer is 
\begin{equation}
	\ts{complete} \approx \ts{form} + \ts{fall}.
\end{equation}
We can then approximate the steady state value of $\pcomp$ as 
\begin{equation}
	\pcomp \approx \frac{1/\ts{complete}}{1/\ts{complete} + 1/\ts{uniform}}. 
\end{equation}

With no free fitting parameters, this simple model -- which accounts only for the production rate of the end and uniform components of the product distribution -- captures remarkably well the range of $\G{end}$ over which $\pcomp$ is high and the maximum value of $\pcomp$ over the whole range of $\G{end}$, as seen in Fig.~\ref{fig:Gend} and Fig.~S5. This fact suggests that the physics incorporated into this simple model explains the majority of the behaviour.  

Specifically, in Fig.~\ref{fig:Gend}, at $\G{end} = 0$, there is no bias for completed polymers to fall off the template. Products are dominated by dissociation from the body of the template, and we see a roughly uniform product length distribution as encountered in Section \ref{subsec:generic} and $\pcomp \approx 1/L_T$. At small values of $\G{end}$, breaking the copy-template bond at the end of the template remains the rate limiting step in producing a full-length copy and $\ts{complete} \approx \ts{end}$. As $\G{end}$ increases, the time to break the end copy-template bond, $\ts{end}$, decreases exponentially and therefore $\pcomp$ increases exponentially. Eventually, however, $\pcomp$  saturates when $\ts{end} < \ts{disp}$; at this point, $\ts{complete} \approx \ts{disp}$ and breaking the copy-template bond at the end is no longer rate-limiting. Here the model suggests that no further increase in $\pcomp$ can be achieved by increasing $\G{end}$; the value of the saturating level is determined by the relative timescales of displacement $\ts{disp}$ and spontaneous detachment from the body of the template $\ts{uniform}$. 

At very high values of $\G{end}$, the simple model accurately predicts the drop in $\pcomp$ to zero shown in Fig.~\ref{fig:Gend}. Here, the end bond is so unstable that the monomer occupancy of the final site on the template, $g$, tends to zero. Consequently, $\ts{form}$ diverges and no full length copies form. Instead the system effectively operates as a uniform template of length $L_T-1$, and generates a uniform distribution of polymers with lengths $L \leq L_T - 1$, but no polymers with length $L = L_T$. This behaviour is particularly clear in the full length distribution shown in Fig.~\ref{fig:GendDist}, in which at high $\G{end}$ no polymers are produced with $L=L_T=10$.

The simple model presented here assumes that the system operates in the dense, ordered brush regime. However, in Fig.~\ref{fig:Gend}, where $\M = 1$ and $\G{gen} = -8$, the constraint given in Eq.~\ref{eqn:constrdense} which predicts whether the dense conformation is reached, $\ts{\rm disp}  L_{T}/2 < \ts{\rm unbind}  /L_T$, is not satisfied. Here the model's estimates of the level of the plateau in $p_{\rm complete}$ are less accurate. The biggest failure of the simple model, however, is not predicting the dips in $p_{\rm complete}$ that occur at intermediate values of $\G{end}$. In Fig.~\ref{fig:Gend}, a single dip in $\pcomp$ is observed for $\M = 1$, and a dip and an intermediate plateau are observed for $\M = 10$. From looking at the full polymer length distributions shown in Fig.~\ref{fig:GendDist} and Figs.~S2, S3 and S4, we can conclude that these dips correspond to a large spike in dimer production. Moreover, from snapshots of simulations we observe that this dimer formation occurs at the final two sites on the template underneath the fraying end of completed- but not yet detached- polymers (Fig.~\ref{fig:GendConfDimer} \textbf{E}).

Despite significant effort, a simple, quantitative model that predicts these `dimer dips' could not be constructed, however a qualitative understanding of the factors which cause dimer production has been reached. The production of dimers occurs at fairly high values of $\G{end}$, which are certainly not necessary for this the copying to be effective (it would be sufficient to just weaken the connection by a few units of thermal energy) and which may be experimentally inaccessible. Nonetheless, we offer an explanation of how this unexpected phenomenon arises in the model. 

When $\G{end}$ is weakened sufficiently that $\ts{end}<\ts{disp}$ (shown as a red line in Fig.~\ref{fig:Gend}), the leading edge of completed polymers are released faster than the tails are fully displaced, leading to less dense conformations such as Fig.~\ref{fig:GendConfDimer} \textbf{B} and \textbf{C}. Completed polymers tend to be displaced from the template from their lagging edge, and can spontaneously fray from their leading edge at the end of the template. When $\G{end} > -\G{gen}-\G{spec} + \ln\Meff$ (equality at green line in Fig.~\ref{fig:Gend}), frayed configurations at the leading edge, such as Fig.~\ref{fig:GendConfDimer} \textbf{c}, are more likely than un-frayed alternatives such as Fig.~\ref{fig:GendConfDimer} \textbf{B}. This constraint on $\G{end}$ locates the dimer dip. If a completed polymer frays far enough to expose the penultimate site on the template, a monomer may occasionally bind there with a strong combined generic and specific bond. This penultimate monomer is unable to polymerise into the completed chain, and instead dimerises with monomers which occasionally bind to the final template site. Completed polymers can be understood to create a 'protective niche' for the formation of dimers. Upon dimerising, the dimer is rapidly released. This cycle is shown in Fig.~\ref{fig:GendConfDimer} \textbf{E}.

If the timescale of formation and release of dimers in the protective niche is less than the average timescale on which the completed polymers are released, then multiple dimers may be released for each completed polymer. When $\G{end}$ is increased further, the monomer occupancy of the final site reduces and the time to form a dimer increases. If the timescale of formation and release of dimers in the protective niche is greater than the timescale on which the completed polymers are released, then the completed polymers fall off, removing the protective niche. As in Fig.~\ref{fig:GendConfDimer} \textbf{E}, without the protective niche, then the dimer is able to polymerise into the preceding polymer, which dramatically reduces the number of dimers that are released into the product pool. Hence, after the dimer dip, we observe a sharp increase in $\pcomp$ again, close to the saturating value that the simple model predicts, before the eventual drop off to $p_{\rm complete} \approx 0$ at the highest values of $\G{end}$. 
At high monomer concentrations, such as $\M = 10$ in Fig.~\ref{fig:Gend}, we observe an intermediate plateau in $\pcomp$ as well as the dimer dip, which occurs when $\ts{end}<\ts{disp}$ shortly after the model predicts saturation. Due to the increased monomer concentration, fraying events at the end of the template result in more monomer invasions at the penultimate site causing more frequent formation of dimers. 
\newpage
\subsection{On-rate discrimination enables the reliable production of copies with the correct length and accurate sequences}
\label{subsec:accuracy}
  
\begin{figure}[b]
	\centering
	\includegraphics[width=\columnwidth]{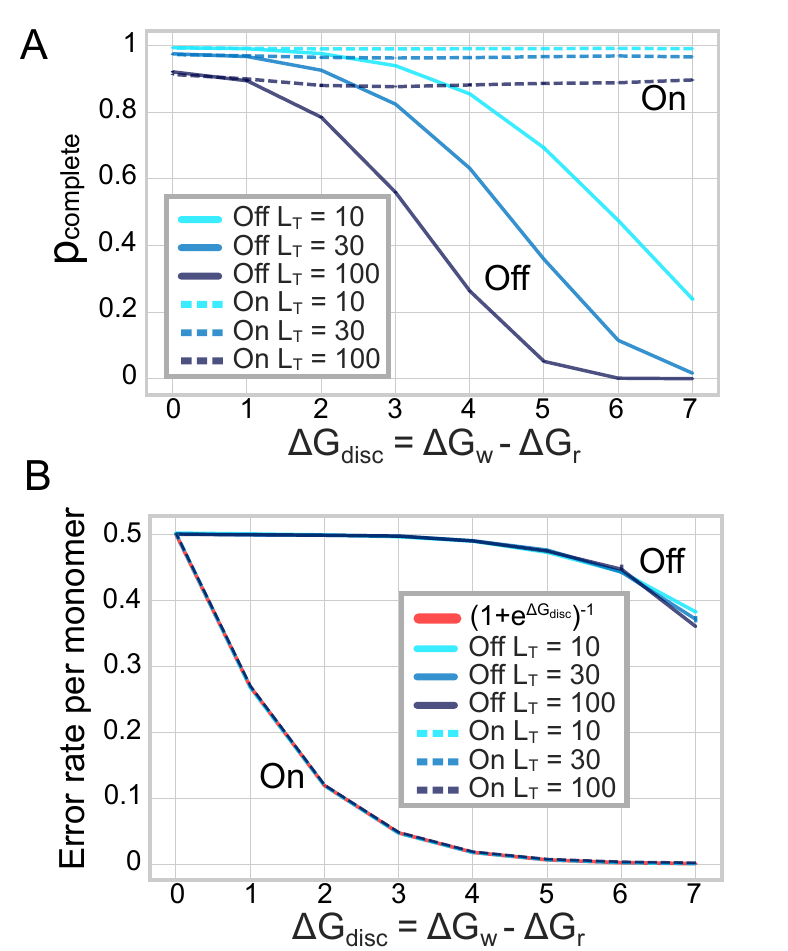}
	\caption{Models with on-rate discrimination can yield complete polymers with high probability, $\pcomp \approx 1$,and  low error rates per monomer, whereas models with off-rate discrimination cannot reliably generate complete polymers with low error rates per monomer. We present \textbf{A} the mean of $\pcomp$ and \textbf{B} the mean error rate per monomer over 5 repeats at each value of the discrimination free energy, $\G{disc} = \G{w} - \G{r}$, for simulations using the model in Fig.~S6 of the Supplementary Material with an alphabet size of two. Simulations are performed for $L_{T} =$ 10, 30 and 100 using $k = 1$, $\G{BB} = -20$, $\G{gen} = -12$, $\G{end} = 7$, and $[M_0]=[M_1]=1$ for models with on- and off-rate discrimination, shown with dotted and solid lines, respectively.  
	Standard Error of the Mean (SEM) is shown as errorbars, which are generally smaller than line width. In \textbf{B} the error rates per monomer for different template lengths overlap one another and lie very close to the limiting line $(1+\exp(\G{disc}))^{-1}$. }
	\label{fig:OnVOff}
\end{figure}

In Section~\ref{subsec:weakend}, we identified a mechanism for producing polymers of a single monomer type with a specific template-determined length in a reliable fashion. Here we consider whether the mechanisms that grant control over the length distribution are compatible with mechanisms which generate accurate copies from a pool of two types of monomer. We extend the models presented in previous sections and consider the growth of binary copies. The parameterisation is given in Fig.~S6 in the Supplementary Material. It is identical to that in Fig.~\ref{fig:EndModel}, except that: (i) $\G{spec}$ now depends on the match between template and copy monomer; (ii) we allow the specific binding free energy to appear either in the binding or the unbinding rates.

In general we now have four specific bond strengths, $\G{00}$, $\G{01}$, $\G{10}$, and $\G{11}$, where the first and second indices give the copy unit and template unit respectively. We make the simplifying assumption that all copy monomer concentrations are equal $[\textrm{M}_0] = [\textrm{M}_1] = \M$ and that all interactions are symmetric (all mismatches are equal, as are all matches) following \cite{Poulton2019}. If the correct pairings, $00$ and $11$, have the same free energy $\G{00}=\G{11}=\Delta G_{r}$, and if the incorrect pairings are also equal, $\G{01}=\G{10}=\Delta G_{w}$, we can describe the specific bond with two energy parameters, $\G{spec,r} = \Delta G_{r}$ or $\G{spec,w} = \Delta G_{w}$, the standard free energy of forming a right match or wrong match copy template bond. Given these simplifying assumptions, we can describe the process of copy sequence simply in terms of whether the monomers are matched (right, r) or unmatched (wrong, w) with respect to the template, allowing us to solve the problem in a way that is independent of the specific template sequence in question. 

Again, we use the principle of local detailed balance constrain the forwards and backwards transition rates between any pair of states. The ratio of the monomer binding rates $R_{\textrm{bind,r/w}}$ to the monomer unbinding rates $R_{\textrm{unbind,r/w}}$ for right or wrong template matches is proportional to the exponential of the specific bond strength (amongst other factors in Fig.~\ref{fig:EndModel}):
\begin{equation}
	\frac{R_{\textrm{bind,r/w}}}{R_{\textrm{unbind,r/w}}} \propto e^{- \G{r/w}}.
\end{equation}

We compare the effects of two contrasting parameterisations of the model, detailed in Fig.~S6 of the Supplementary Material. If the unbinding rates are dependent on the match between copy and template, $R_{\textrm{unbind,r/w}} \propto e^{\G{r/w}}$, as was the case in the models presented in previous sections, then the system has "off-rate" discrimination, and the monomer binding rates are constant. If the monomer binding rates are dependent on the match between copy and template, $R_{\textrm{bind,r/w}} \propto e^{-\G{r/w}}$, then the system is has "on-rate" discrimination, and the unbinding rates are constant. As shown in Fig.~S6 of the Supplementary Material, we apply the $\G{spec}$ sensitivity consistently across all binding and unbinding transitions for on and off-rate discrimination. As in the Section~\ref{subsec:weakend}, all unbinding rates for units at the end of the template are increased by a factor $e^{\G{end}}$.

We take parameter values that produced complete polymers in Section \ref{subsec:weakend}. To generate the data shown in Fig.~\ref{fig:OnVOff}, we set $k = 1$, $\G{BB} = -20$, $\G{gen} = -12$, $\G{end} = 7$, and $[M_0]=[M_1]=1$. We set $\G{spec,w} = \Delta G_{w} = -1$ for wrong-match monomers, and $\G{spec,r} = \Delta G_{r} = \{-1,-2,\ldots,-8\}$ for right-match monomers. 
As described in the methods section, to generate each data point in Fig.~\ref{fig:OnVOff} we ran 5 independent repeats of simulations that each generate up to 5000 copolymers. In addition to calculating the average of $\pcomp$, we also calculate the average error rate per monomer (the number of wrong units in the copolymer divided by the polymer length) for complete polymers. 

In Fig.~\ref{fig:OnVOff} \textbf{A}, it is evident that on-rate discrimination maintains a high proportion of complete polymers even at long lengths and large values of $\G{disc} = \Delta G_w - \Delta G_r$. Off-rate discrimination, however fails to produce long polymers as the discrimination free-energy difference, $\G{disc} = \G{0} - \G{1}$, grows larger. The generic bond and weak end mechanisms presented in the previous sections, which generate and selectively release long polymers, require the leading edge of polymers to remain bound to the template until the end of the template is reached. However, if the system discriminates on the off-rates, then the incorporation of incorrect monomers into a polymer can cause it to be released prematurely, thereby reducing $\pcomp$.

In Fig.~\ref{fig:OnVOff} \textbf{B}, on-rate discrimination delivers low error rates per monomer, close to the equilibrium ratio implied by the discrimination free energy $1/(1+\exp(\G{disc})$), even though the products are far from equilibrium after separation from the template. Off-rate discrimination is inaccurate for moderate discrimination energies as a strong $\G{gen}$ allows incorrect monomers to be incorporated into copolymers. For off-rate discrimination, there is a trade-off between accuracy and $\pcomp$ for the mechanisms we have considered here. Hence, on rate discrimination is needed to gain control of the length and accuracy in these minimal models of autonomous templated copying. 

\section{Conclusion}

\label{sec:conc}
Unlike the template-copying machinery found in biological systems that can cyclically copy templates under constant environmental conditions \cite{Reha-Krantz2010,Joyce2004,Vannini2013,Ebright2000,Hirose2000,Rodnina2018,Fox2010,Petrov2015,Bowman2020,Nikolay2015,Caetano-Anolles2015}, state-of-the-art synthetic molecular copying systems require external intervention or non-chemical driving \cite{He2017,Zhuo2019,Nunez-Villanueva2019,Schulman2012, Braun2004,Mast2010,Mast2013,Kreysing2015} to operate cyclically. Here we have developed coarse-grained models of isothermal, enzyme-free, templated copolymerisation processes in which the copy copolymers can spontaneously detach from the template in a constant environment, driven by chemical free energy alone.

The central challenge of producing polymer copies is overcoming product inhibition, which is exacerbated by the natural cooperativity of the interactions between two polymers. We have shown that a surprisingly simple mechanism -- using the free-energy of polymerisation to weaken the interaction between the monomers and the template behind the leading edge of the copy -- is sufficient to generate long polymers reliably.  Combining this mechanism with a weakened copy-template bond at final site on the template, as in Section \ref{subsec:weakend}, is sufficient to reliably generate copies of the full template length under a wide range of conditions. 

These proposed mechanisms require: 1) strong template-monomer bonds (so that the leading edge of growing polymers remain bound to the template for long enough to extend along its full length); 2) strong backbone bonds between monomeric units within polymers (so that formed polymers don't spontaneously fragment) that, upon formation 3) cause major disruption to the copy-template bond of the 'back' monomer (so that the long cooperatively binding tails of polymers can be displaced from template by other polymers); and 4) a moderate destabilisation of the copy-template bond at the end of the template (to enable the selective release of fully-formed polymers). To copy specific sequences in addition to producing polymers of a fixed length, it is necessary to also have ``on-rate" discrimination, in which matching sequences bind to the template faster. 

\markup{ The conditions stated above can be met by DNA-based chemical reaction systems that exploit handhold-mediated DNA strand displacement (HMSD) \cite{Cabello-Garcia2021}. In HMSD, polymerisation between monomers on the template results in the weakening of the copy-template connection in the `lagging' monomer through the transfer of bonds via strand displacement. Moreover, templating via HMSD has the potential to achieve effective on-rate discrimination during monomer binding, using a mechanism in which a transient match-specific toehold forms first, with the strong generic bond only forming after an additional (toehold-mediated) strand displacement reaction\cite{Zhang2009,Srinivas2013}.}

The models we have introduced have a surprisingly rich range of behaviours and their dynamics did not match our initial expectations. We expected that optimal copying would involve isolated copies on a template, as is frequently observed in transcription \cite{Vannini2013,Ebright2000,Hirose2000} and translation \cite{Rodnina2018,Fox2010,Petrov2015,Bowman2020,Nikolay2015,Caetano-Anolles2015}, and as reflected in previous models of templated copolymerisation \cite{Poulton2019}. In such a setting it would be necessary to prohibit copies from starting at random locations on the template, and thereby generating truncated products. However, the natural tendency of brush-like configurations observed here to absorb isolated monomers into growing copies at all points on the template solves that challenge, which otherwise might require complex activation and deactivation of template sites ahead of and behind the growing copy. Moreover, the observed displacement behaviour means that copying is possible even when the tails of polymers have moderate affinity with the template; indeed, this regime was more effective in producing longer polymers than regimes in which tail affinity was so low that they would detach without displacement, as described in Section S~3 and Fig.~S5 of the Supplementary Material.

Extant biological template copying processes, such as transcription \cite{Vannini2013,Ebright2000,Hirose2000} and translation \cite{Rodnina2018,Fox2010,Petrov2015,Bowman2020,Nikolay2015,Caetano-Anolles2015}, can operate in sparse regimes in which the copies do not fully occupy the template, although reasonably dense regimes are also observed\cite{Nanikashvili2019,Zur2016}. The sparse regime of operation is possible because the initiation of polymer growth only occurs at specific locations on the template, at ribosome binding sites in translation for instance \cite{Rodnina2018}, in tightly controlled ways that involve complex enzymatic machinery. Similarly, the enzymatic machinery of transcription and translation provides a larger, moving window in which copies are bound to their template \cite{Rodnina2018,Fox2010,Petrov2015,Bowman2020,Nikolay2015,Caetano-Anolles2015}. This factor enables off-rate discrimination between the incoming monomers without causing partially formed polymers to fall off prematurely. It is challenging see how to engineer this in a simple setting without the interplay of enzymes that can define a scale for this template-attached region, but such a scheme may facilitate additional mechanisms such as kinetic proofreading to increase the accuracy above limit set by the free energetic discrimination between monomer types. 

\markup{ Mechanisms that rely on displacement of copies by the subsequent copies have been considered as solutions to product inhibition in an RNA world \cite{Tupper2021,ZhouSzostak2019}. In particular, Tupper and Higgs have argued that a rolling-circle copying of a circular template can alleviate product inhibition via strand displacement \cite{Tupper2021}. This rolling circle approach eliminates the tendency of short, newly-initialised copies to be out-competed by longer polymers on a linear template, as we observed in Section~\ref{subsec:basic}. The mechanism assumes directional polymerisation, but polymerisation is still potentially hampered by a displacement-based competition between  polymer ends for binding to the template, with no intrinsic bias favouring the desired pre-polymerisation configuration. 
}
\markup{
The asymmetric destabilisation in our model induces a directionally-biased strand displacement between copy polymers, which can overcome product inhibition even on linear templates. Our mechanism may also work well in conjunction with the rolling circle mechanism of Tupper and Higgs, providing a directional bias to resolve the competition for the template. However, the chemistry and bonding stability of individual nucleotides as explored by Tupper and Higgs \cite{Tupper2021} -- as opposed to larger oligonucleotide monomers as used in processes like HMSD \cite{Cabello-Garcia2021} -- may not be well-suited to the mechanisms we have investigated in this work. Outside of DNA-based reactions, Osuna Gálvez and Bode's reaction -- in which polymerisation is directly coupled to the disruption of product-template connections -- shares the most resemblance to our mechanisms, though the templates cannot be reused. It is an open question whether small organic molecules with the appropriate chemistry can be identified and used to build synthetic copying systems with the properties we've stipulated.}

We have attempted to keep the models as simple and general as possible, so that they can be used to guide the design of synthetic copying systems in a wide array of contexts. We have also attempted to make all chemical steps explicit, rather than invoking a chemical {\it deus ex machina} that resolves the central challenges of the copy processes, just as highly-evolved enzymatic machinery do {\it in vivo}. A key assumption of our model, however, is that (de)polymerisation only occurs while monomers are attached to the template - the template acts as a catalyst for bond formation and breakage. A passive template that simply brings reactants into close proximity could certainly allow polymerisation reactions that are prohibitively slow in solution to proceed on the template. However, such a mechanism would not accelerate depolymerisation as well, and so unless the template has some direct chemical coupling to the bond formation mechanism, as is typical in enzymes \cite{Reha-Krantz2010,Joyce2004,Vannini2013,Ebright2000,Hirose2000,Rodnina2018,Fox2010,Petrov2015,Bowman2020,Nikolay2015,Caetano-Anolles2015}, it is hard to see how to justify this catalytic assumption. Indeed, as we saw in Section \ref{subsec:basic}, simple templates are not effective catalysts in our model -- they accelerate the polymerisation reaction by forming a stable complex with the product, preventing cyclic copying. However, in our modified mechanism presented in Section \ref{subsec:generic}, the competition between the polymerisation bond and the generic bond with the template provides a direct mechanism for catalysis, justifying the assumption that (de)polymerisation only occurs while monomers are attached to the template.

We have also assumed that polymers that detach from the template and are released into the bath do not rebind to the template. One could imagine, if the concentration of polymers in the bath grew to significant levels, that rebinding could occur. Indeed, the rebinding and subsequent elongation of products was presented as a plausible pathway to the generation of longer RNA oligomers, bridging a portion of the gap between short RNA building blocks and the long RNA molecules that make up ribozymes in a pair of papers on templated ligation in an RNA world \cite{Rosenberger2021,Kudella2021}. However, in our  mechanism, polymerisation is coupled to the disruption of the copy-template bonds,  reducing the interaction strength that most units in the polymer chain could rebind by. Only the leading edge of a grown polymer could potentially stick tightly to the template, possibly causing unwanted over-extension of polymers if this reattachment happened at an earlier site on the template that happened to be complementary. If one were to design a copying machine based upon the biased displacement mechanism introduced here, a special monomeric unit could be reserved to bind effectively only to the ends of templates, and thus cap the end of completed polymers. This approach would prevent further unwanted extension and reduce the copy-template affinity for completed products, alleviating the aforementioned issues. 

Another key assumption of this work is that there is no interaction between polymer tails. Our models suggest a dense brush of polymers is beneficial for length control, however this may introduce interactions between polymer tails, especially on long templates, in {\it in vitro} settings. The importance of the interaction between polymer tails and the formation of bulges on the template should be considered on a case-by-case basis in experimental settings.

Finally, in our results we have not focused on the speed with which copies are produced, only the relative yield of longer or full-length copies. In this setting we observed that polymer tails that have some affinity to the template are advantageous in maximising the likelihood of producing full length copies. However, such this behaviour likely comes at the cost of somewhat slowed copy production, since template sites are more likely to be blocked. The optimal balance will depend on the system in question, and again will require further investigation on a case-by-case basis.   

\section{Data Availability}
\label{dataavail}
The code, input files, and analysis scripts required to reproduce the data presented in this work are available for download from the Zenodo file repository through the following link:  \href{https://doi.org/10.5281/zenodo.5643000}{DOI 10.5281/zenodo.5643001}.
\section{Supplementary Material}
The Supplementary Material contains a description of parameter values excluded from Fig.~\ref{fig:SimpleModelSweep} which generated few samples, the full polymer length distributions accompanying Fig.~\ref{fig:Gend}, a demonstration that the asymmetric destabilisation mechanism is more effective when polymer tails have higher affinity for the template, and a depiction of the model used in Section~\ref{subsec:accuracy}, for which there are two types of monomer.

\section{Author Contributions}
All authors conceived the project. JJ wrote the simulations and performed the analysis. All authors interpreted the results and co-wrote the paper.
\section{Conflict of Interest}
The authors declare no conflicts of interest.
\begin{acknowledgments}
This work is part of a project supported by the European Research Council (ERC) under the European Union's Horizon 2020 research and innovation programme (Grant agreement No. 851910). JJ is supported by a Royal Society studentship. TEO is supported by a Royal Society University Research Fellowship.
\end{acknowledgments}

\bibliographystyle{jcp.bst}
\bibliography{biblio.bib}

\FloatBarrier
\end{document}


\preprint{AIP/123-QED}
\title{Supplementary Information for "Minimal mechanism for cyclic templating of length-controlled copolymers under isothermal conditions"}

\beginsupplement
\maketitle
\section{Extreme parameter values create long embedded processes}
\label{sec:longembedded}
\begin{figure}[h]
	\centering
	\includegraphics[width=\textwidth]{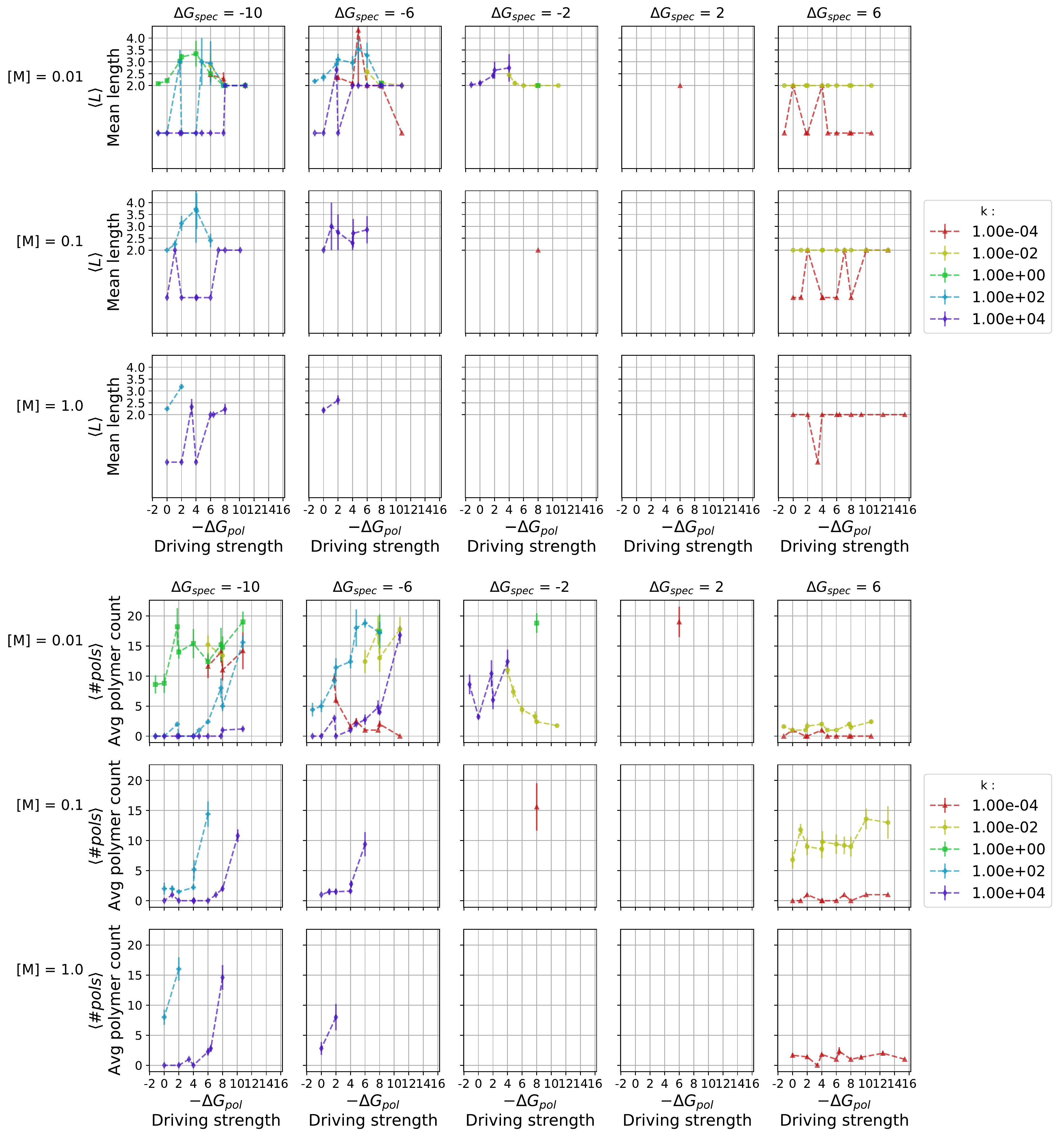}
	\caption{ Extreme parameter values result in long embedded processes which are time consuming to simulate with the Gillespie algorithm \cite{Gillespie1976}. These plots are the average length and average polymer count produced by simulations with average polymer count below 20. When $\G{spec}$ is high, slow polymerisation rates $k$ lead to undersampling, and when $\G{spec}$ is low, fast polymerisation rates and low monomer concentrations $\M$ lead to undersampling. Undersampling is due to long embedded processes. }
	\label{fig:LowSampling}
\end{figure}

In Fig.~\ref{fig:LowSampling}, we show that extreme parameter values result in long embedded Markov processes which are time consuming to simulate with the Gillespie algorithm \cite{Gillespie1976}, resulting in simulations that produce few polymers. We plot the averaged mean polymer length and averaged polymer count produced by parameter values which generated fewer than 20 polymers in Fig.~\ref{fig:LowSampling}. When $\G{spec} = 6$, the copy-template interaction is very weak, and the dynamics are dominated by repeated binding and unbinding of monomers to the template. Hence when the polymerisation rate is also slow, $k = 10^{-4}$ and $k = 10^{-2}$, many repeated binding and unbinding events occur before the formation of a dimer, and very few polymers were produced by the simulation within the run time. When $\G{spec} = -10 $, the copy-template interaction is very strong, unbinding is slow. Hence, for fast polymerisation rates $k$ and low values $-\G{pol}$, the dynamics are dominated by repeated polymerisation and depolymerisation of neighbouring units, and it takes a long time to simulate the formation and release of dimers. In cases where few outputs of polymers were given, it doesn't necessarily give long polymers, but we have no evidence that there is a tendency to produce long polymers reliably.

\FloatBarrier
\subsection{Full product length distributions accompanying main text Fig.~12}

\begin{figure}[h]
	\centering
	\includegraphics[width=\textwidth]{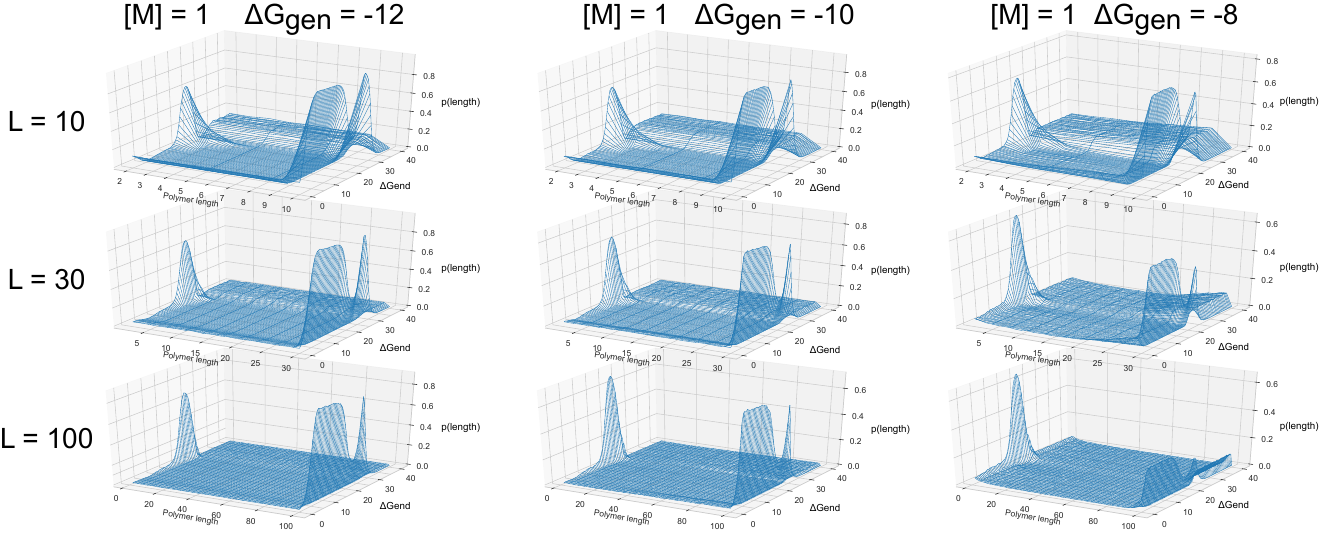}
	\caption{ The product probability length distributions $p(\textrm{length})$ averaged over 5 independent repeats of ensembles of up to 1000 polymers at each value of $\G{end} = 0, 0.5, ... , 40$, for $\M = 1$, $\G{gen} =\{-12,-10,-8\}$, and $L_{T} = \{10,30,100\}$.}
	\label{fig:GendDists1}
\end{figure}

\begin{figure}[h]
	\centering
	\includegraphics[width=\textwidth]{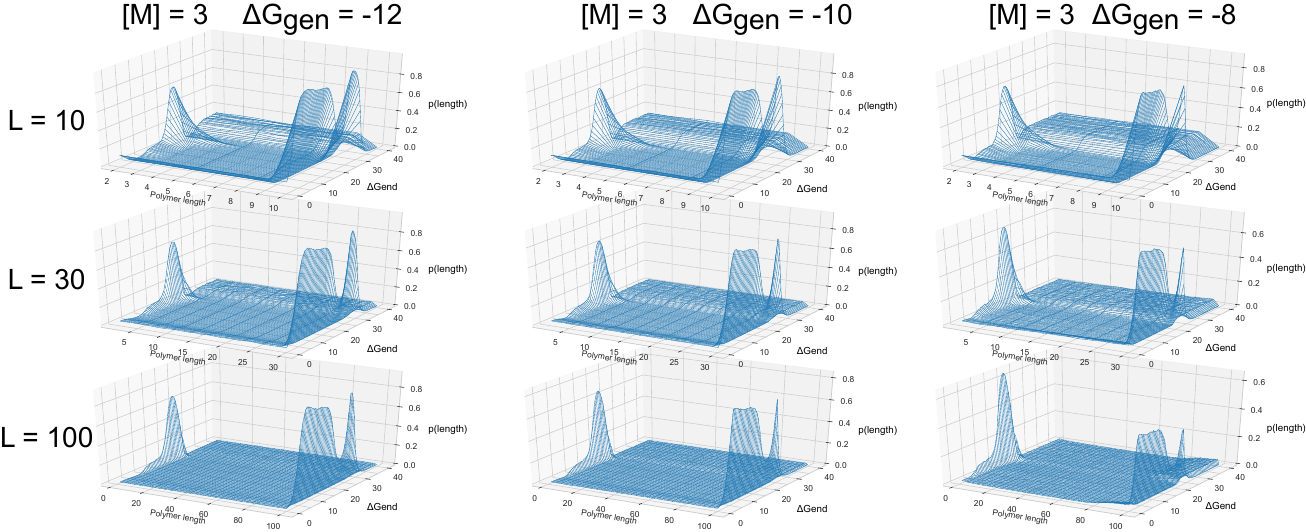}
	\caption{ The product probability length distributions $p(\textrm{length})$ averaged over 5 independent repeats of ensembles of up to 1000 polymers at each value of $\G{end} = 0, 0.5, ... , 40$, for $\M = 3$, $\G{gen} =\{-12,-10,-8\}$, and $L_{T} = \{10,30,100\}$. }
	\label{fig:GendDists3}
\end{figure}

\begin{figure}[h]
	\centering
	\includegraphics[width=\textwidth]{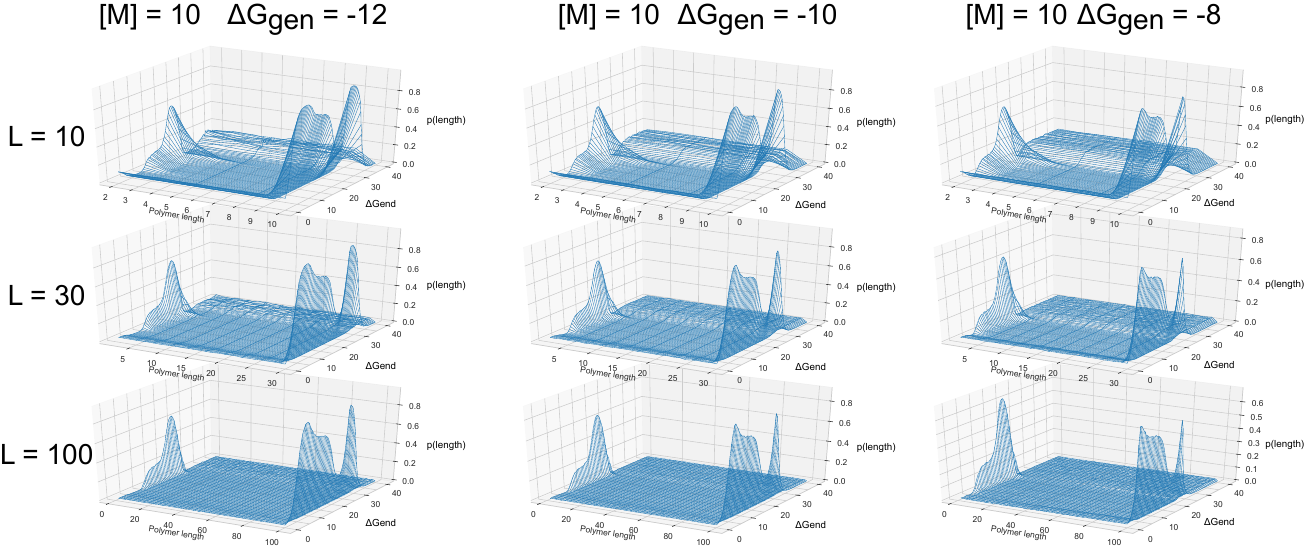}
	\caption{ The product probability length distributions $p(\textrm{length})$ averaged over 5 independent repeats of ensembles of up to 1000 polymers at each value of $\G{end} = 0, 0.5, ... , 40$, for $\M = 10$, $\G{gen} =\{-12,-10,-8\}$, and $L_{T} = \{10,30,100\}$. }
	\label{fig:GendDists10}
\end{figure}

In Section~III.C of the main text, we demonstrated that weakening the copy-template interaction strength at the final site of the template with an energetic factor $\G{end}$ was sufficient to selectively release complete polymers. In Fig.~12, we presented the only the complete-length component, $\pcomp$, of the product length distribution as $\G{end}$ was varied. In Figs.~\ref{fig:GendDists1}, \ref{fig:GendDists3} and \ref{fig:GendDists10} we show the entire product length probability distribution, $p(\textrm{length})$ against $\G{end}$, for all values of $\M = \{1,3,10\}$, $\G{gen} = \{-12,-10,-8\}$, and $L_{T} = \{10,30,100\}$. We fix $k = 1$, $\G{spec} = -4$ , and $\G{BB} = -20$, which puts the system into a dense brush regime in which long copies with $\langle L \rangle \sim L_T/2$ are produced in Section~III.B of the main text when $\G{end} = 0$. $\pcomp$ corresponds to the slice of the length distribution for polymers which have the same length as the template $p(\textrm{length} = L_{T}) \equiv \pcomp$.

As discussed in Section~III.C of the main text,, when $\G{end} = 0$, in the surface plots below we see a relatively flat, uniform length distribution. When is increased $\G{end}$ to moderate values complete polymers dominate the length distribution. From the length distributions shown below we can clearly see that the dip in $\pcomp$ that features in Fig.~12 is due to a peak in the production of short polymers. At the highest values of $\G{end}$, no complete polymers are produced ($p(\textrm{length} = L_{T}) \equiv \pcomp \approx 0$) and the rest of the length distribution is flat. 


\FloatBarrier
\clearpage
\subsection{The asymmetric destabilisation and weak end mechanism is more reliable for polymer tails with higher template affinity}
\label{sec:SIGspec}
\begin{figure}[b]
	\centering
	\includegraphics[width=0.8\textwidth]{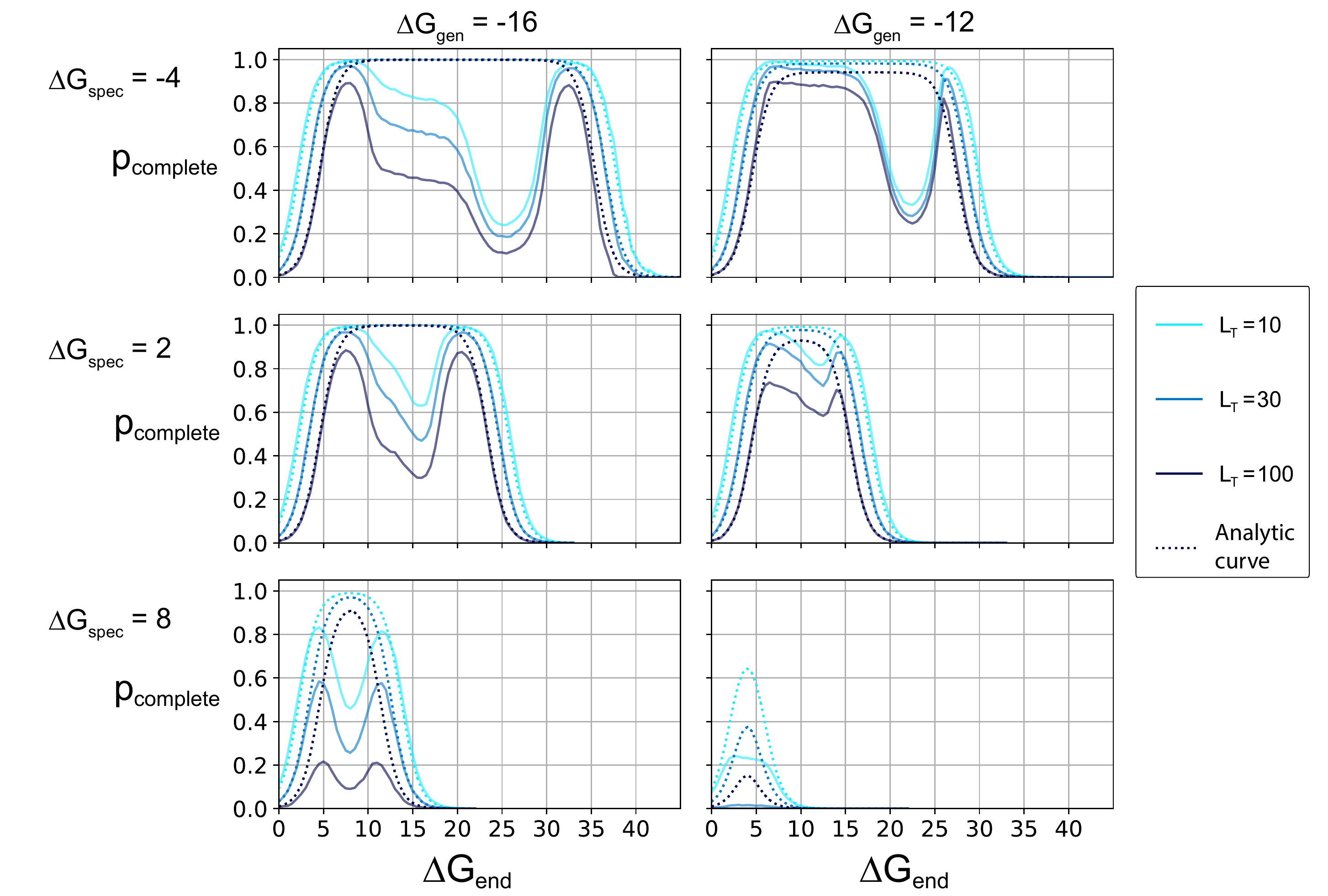}
	\caption{ When the specific bond strength, $\G{spec}$, is weakened/increased, the range of $\G{end}$ over which $\pcomp$ is increased is reduced. The simple analytical model (dotted line) presented in Section~III.C of the main text still reproduces the boundaries over which $\pcomp$ is high with no free fitted parameters. The simple model excludes the effects of dimer formation that occur at intermediate values of $\G{end}$, which makes $\pcomp$ lower than the predicted value. }
	\label{fig:GendGspecGgen}
\end{figure}

In addition to the results presented in Section~III.C of the main text, in Fig.~\ref{fig:GendGspecGgen} we demonstrate that the combination of the asymmetric destabilisation of polymer tails induced by $\G{gen}$ and the weakening of the end of the template with $\G{end}$ can result in the reliable production of complete polymers $\pcomp$ even when the tails of polymers tend to detach from the template faster than they rebind with $k_0 exp(\G{spec}) \gg k_0 \Meff$, or when $\ln{\G{spec}} > \ln{\Meff}\approx 4.6$. Here we set $k_0 = k = 1 $, $\M = 1, [\Meff] = 100 , \G{BB} = -20, \G{gen} = \{-16,-12\}, \G{spec} = \{-4,2,8\}$ and $\G{end} = \{0,0.5,...\}$ and $L_{T} = \{10,30,100\}$, and calculate the average of $\pcomp$ from up to 2000 polymers produced by 5 independent simulations at each point in the parameter space. 

In Fig.~\ref{fig:GendGspecGgen}, as $\G{spec}$ is increased, $\pcomp$ begins to fall at lower values of $\G{end}$. This cusp in $\pcomp$ is well reproduced by the simple model presented in Section~III.C of the main text. As $\G{end}$ increases, the occupancy of the final template site reduces, which increases the average timescale on which polymers are formed (completed) $\ts{form}$. When the timescale to form a complete polymer $\ts{form}$ is of the same order as the timescale on which the $L_{T}$ incomplete polymers detach from the template $\ts{uniform}$, the flux of uniform length polymers exceeds the flux of completed polymers, and hence $\pcomp$ drops. This drop, which defines the useful range of $\G{end}$, occurs at lower values of $\G{end}$ when the specific bond strength $\G{spec}$ is weaker (more positive). 

Additionally in Fig.~\ref{fig:GendGspecGgen},  we note that for $\G{spec} = 8$, $\pcomp$ fails to increase substantially upon weakening the end of the template, especially for long templates. We note that the for $\G{gen} = -16$, at $\G{end} = 0$, the system lies in region \textbf{III'} of Fig.~7 of the main text, where a dense brush of displaced polymer tails is expected to form on the template, which results in a uniform product length distribution. However, for $\G{gen} = -12$, at $\G{end} = 0$, the system lies in region \textbf{V} where the spontaneous detachment of polymers enables fragments to grow, biasing the production of shorter polymers, as the system doesn't reach a maximally-displaced conformation. Hence, for $\G{gen} = -16$, $\G{spec} = 8$, $\pcomp$ fails to increase substantially upon weakening the end of the template because the workable range of $\G{end}$ is slim. However, for $\G{gen} = -12$, $\G{spec} = 8$, $\pcomp$ is further reduced because of the tendency for the system to release incomplete polymers before reaching a maximally-displaced conformation in which weakening the end of the template can selectively release completed polymers. 

\FloatBarrier
\clearpage
\subsection{Model of templated polymerisation with a binary monomer pool}
\begin{figure}
	\centering
	\includegraphics[width = 0.9\textwidth]{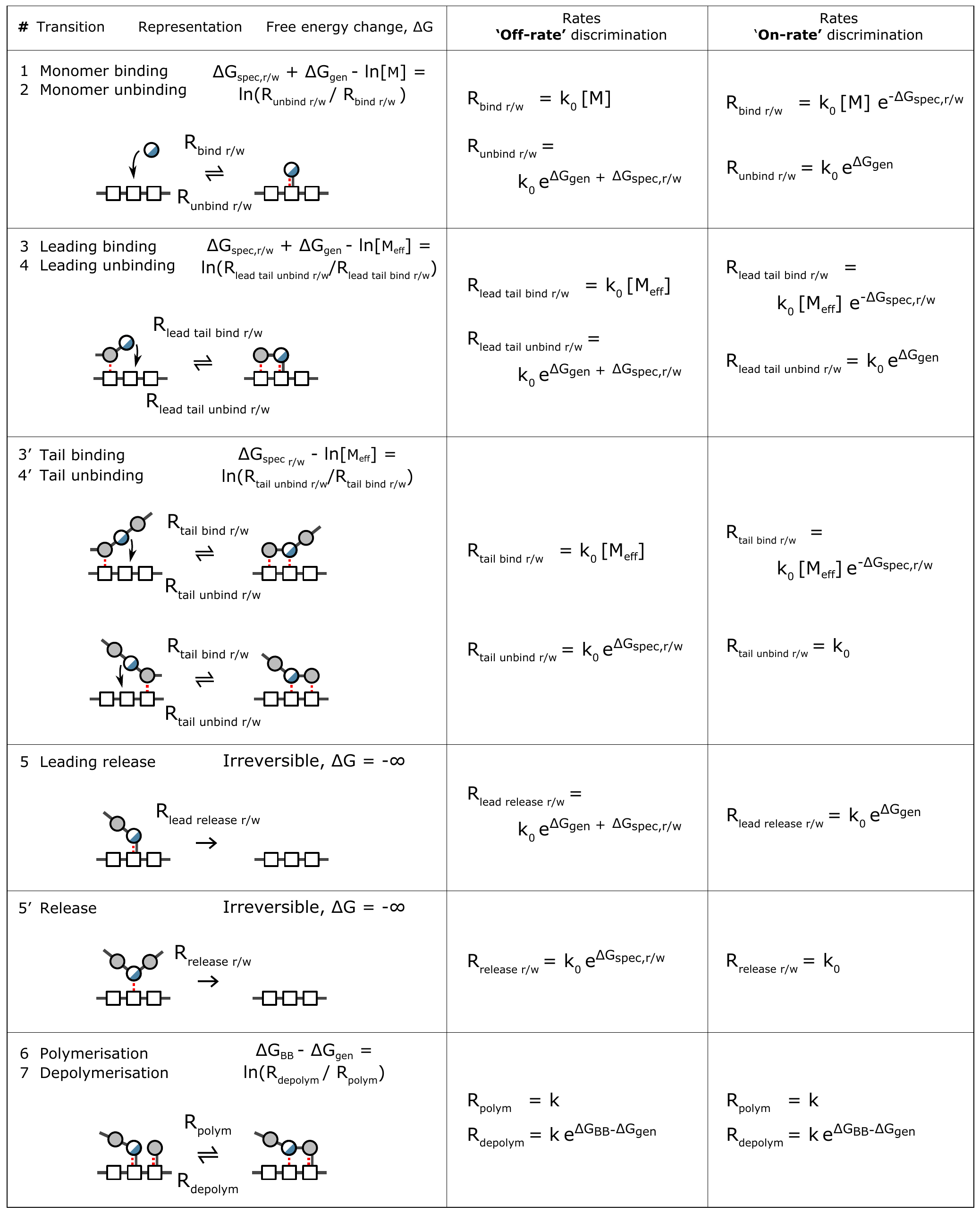}
	\caption{In a system with two competing monomer types, "right" (white) specfic copy-template bonds have a strength $\G{spec,r}$ and "wrong" (blue) specific copy-template bonds have a strength $\G{spec,w}$. We have half-shaded the monomers which affect the transition rates. The identity of monomers shaded gray do not affect the transition rates. We consider parameterisations in which discrimination occurs on the binding step (''on-rate'' discrimination) or on the unbinding step (''off-rate'' discrimination) as described in Section~III.D of the main text. Unbinding reactions at the end of the template are sped up by a factor $exp(\G{end})$ for both monomer types as depicted in Fig.~10 of the main text but are not shown here for brevity.}
	\label{fig:OnOffModel}
\end{figure}

In Fig.~\ref{fig:OnOffModel}, we depict the model used in Section~III.D of the main text where the monomer alphabet size is two. As described in Sec.~III.D of the main text, the monomers form either a 'right' or 'wrong' match with each template site. 'Right' monomers (white) bind with a total bond strength $\G{spec,r} + \G{gen}$ and 'wrong' monomers (blue) with a total bond strength $\G{spec,r} + \G{gen}$, where $\G{spec,r}\le\G{spec,w}$. The transition rates for right and wrong monomer types have the same form, but the $\G{spec}$ factor is replaced with the value of $\G{spec,r}$ or $\G{spec,w}$ for the corresponding monomer type. For instance, then monomer unbinding rate for off-rate discrimination is $R_{\textrm{unbind,r}} \propto exp(\G{spec,r})$ for 'right' monomers and $R_{\textrm{unbind,w}} \propto exp(\G{spec,w})$ for 'wrong' monomers. For on-rate discrimination, $R_{\textrm{bind,r}} \propto exp(\G{spec,r})$ for 'right' monomers and $R_{\textrm{unbind,w}} \propto exp(\G{spec,w})$ for 'wrong' monomers

\FloatBarrier
\bibliographystyle{jcp.bst}
\bibliography{biblio.bib}